\documentclass[11pt]{article}
 
\usepackage{amsmath}
 \usepackage{tabularx}
 \usepackage{array}
 \usepackage{booktabs}
 \usepackage{graphicx}
 \usepackage{longtable}
 \usepackage{amsfonts}
 \usepackage{float}
\usepackage{hyperref}
\usepackage{bm, geometry, cite, amssymb}
\usepackage{subcaption}
\usepackage{dblfloatfix}
\usepackage{cuted}
\usepackage{caption}
\usepackage{tikz}
 
\begin{document}
 
 \title{\LARGE \bf {Evaluating Chern Topology in Discretized Brillouin Zones Using Bargmann Invariants}
 }

\author{
Swarup Sangiri$^{1}$\thanks{swarupsangiri@iitkgp.ac.in, swarup.phys@gmail.com},
and A Taraphder$^{1}$\thanks{arghya@phy.iitkgp.ac.in}
}

\date{}

\maketitle

\begin{center}
\small
$^{1}$ Department of Physics, Indian Institute of Technology Kharagpur, Kharagpur, 721302, India\\
\end{center}

\vspace{0.3cm}

\begin{abstract}
We develop a finite-state geometric framework based on third-order
Bargmann invariants for evaluating Chern numbers and characterizing
local quantum-state geometry. Using gauge-invariant state overlaps, the
framework evaluates the Chern number directly on discretized parameter
spaces while retaining local geometric information through the
phase-derived density and the complementary amplitude density, as well
as signatures associated with changes in topological character. We
apply the construction to the Qi--Wu--Zhang, $1/3$- and $1/4$-flux
Hofstadter, SSH, and Rice--Mele models, including a rank-$2$ extension
for isolated composite subspaces and applications to closed-loop and
parameter-space topology. The resulting framework organizes Chern
topology and local geometric information directly from finite-state
data, providing a compact state-based description without relying on
dynamical evolution.
\end{abstract}
\newpage
\section{Introduction}
\label{sec:introduction}

Geometric phases provide a direct connection between the structure of
quantum states and observable topological phenomena. Since the
formulation of the geometric phase for adiabatically evolving quantum
states~\cite{Berry1984}, geometric descriptions have become an
important component of the study of quantum systems with nontrivial
topological structure. In band theory, the Berry connection and its associated curvature provide a
local description of momentum-space geometry, while their integral over
a closed Brillouin zone gives the Chern number of an isolated band
~\cite{Thouless1982,Kohmoto1985}. This framework underlies the
quantization of Hall response and the modern description of topological
bands, while Berry-phase and Berry-curvature effects also enter
transport responses in topological materials~\cite{NandyTaraphderTewari2018}.
The Hofstadter problem provides a paradigmatic setting in which
magnetic subbands acquire nontrivial topological character through
their momentum-space structure~\cite{Hofstadter1976}. The connection
between bulk topology and physical boundary phenomena has further
established Chern topology as a central organizing principle of
two-dimensional topological phases~\cite{QiWuZhang2006a,QiWuZhang2006b}.
More generally, topological characterization need not be restricted
to an individual band: when bands touch internally, the topology of
their combined spectral subspace remains well defined provided that
the subspace is separated from the rest of the spectrum
~\cite{AvronSeilerSimon1983}. This also motivates the use of
matrix-valued geometric structures for degenerate or composite quantum
subspaces~\cite{WilczekZee1984}.

An alternative description of quantum-state geometry is provided by
Bargmann invariants, which are gauge-invariant products of overlaps
between quantum states~\cite{Bargmann1964}. Their connection with geometric phases provides a formulation in which
geometric information can be expressed directly through finite sets of
quantum states, without requiring derivatives with respect to continuous
parameters~\cite{Mukunda2003}.
The geometry of quantum-state manifolds also contains information in
the magnitudes of state overlaps, in addition to their phases
~\cite{ProvostVallee1980}. These properties make Bargmann invariants
natural building blocks for a finite-state description of quantum
geometry. Recent applications have demonstrated the utility of Bargmann
invariants in geometric and topological problems ranging from
correlated quantum systems to Weyl, Kitaev, and Majorana-related
structures~\cite{SangiriSarkar2023Baryogenesis,
Sangiri2026Bargmann,Sangiri2026Sequential,
SangiriTaraphder2026Weyl,SangiriSarkarTaraphder2026}. In condensed
matter systems, related developments have explored finite-state
geometric constructions, discrete quantum geometry, and
Bargmann-trace-based descriptions of multiband Bloch-state geometry
and topology~\cite{Shashidhara2017,YahyaviHetenyi2017,
AvdoshkinPopov2023,Huang2026,FarragMeleLow2026,
Reascos2023,LiWagnerZhang2026}. These developments motivate a
finite-state treatment of band geometry in which quantum states at
discrete parameter-space points provide the elementary geometric data.

Discrete approaches have provided gauge-invariant formulations for
evaluating Chern numbers on finite Brillouin-zone meshes
~\cite{Fukui2005}. Motivated by these developments, we consider the
quantum states assigned to the mesh points directly and use the
third-order Bargmann invariant formed from the states at the vertices
of an oriented triangle as the elementary geometric object. Its phase
provides a gauge-invariant finite-state geometric phase. Neighboring
triangles can then be combined into finite cells, with the phase
contributions associated with shared internal links cancelling under
the corresponding orientation. This allows the geometric phase
associated with each finite cell, as well as its accumulation over the
closed Brillouin zone, to be constructed directly from the quantum-state
overlaps.

For a spectrally isolated band on an admissible finite triangulation, we
establish the relation between the accumulated Bargmann phase and the
conventional Chern number. The resulting Bargmann-evaluated Chern
number is the conventional Chern number expressed through
gauge-invariant finite-state overlaps. Retaining the phase of each
elementary cell before the global summation further defines the
finite-resolution geometric density $\mathcal K_B$, whose fine-mesh
limit connects to the Berry curvature. The modulus of the same
finite-state overlap products provides the complementary amplitude
density $\mathcal M_B$, retaining information on the variation of
overlap magnitudes across the parameter space. The continuum Berry
connection and curvature therefore serve as analytical references for
the finite-state construction and its continuum correspondence, rather
than as inputs to the Bargmann evaluation itself.

We demonstrate the construction across complementary lattice models.
For the Qi--Wu--Zhang model~\cite{QiWuZhang2006a,QiWuZhang2006b}, the
finite-state Bargmann phase accumulation reproduces the distinct Chern
sectors of the lower band, while the finite-state admissibility
structure tracks the changing overlap geometry near the critical
parameters. We then apply the construction to the $1/3$-flux
Hofstadter model~\cite{Hofstadter1976}, where the middle band is
globally isolated. The finite-state phase sum gives $C_B=-2$ on the
admissible meshes considered, while the cell-resolved $\mathcal K_B$
reveals strongly nonuniform geometric structure across the magnetic
Brillouin zone. The corresponding $\mathcal M_B$ provides complementary
finite-resolution information on the overlap amplitudes, and a
continuum multiband analysis at representative $\mathcal K_B$
extrema relates these features to the underlying interband matrix
elements and energy separations.

The $1/4$-flux Hofstadter model extends the construction to an isolated
rank-$2$ composite subspace containing two internally touching central
bands. The individual central bands are not globally isolated, whereas
the combined subspace remains separated from the outer bands. We
therefore construct the rank-$2$ Bargmann invariant from determinants
of overlap matrices between local frames, obtaining the composite
Chern number $C_{23}=-2$ on the admissible meshes considered. The
corresponding local rank-$2$ phase distribution provides a
momentum-resolved description of the geometric phase accumulated by
the composite subspace.

Finally, the SSH and Rice--Mele models demonstrate how the finite-state
overlap construction extends beyond the two-dimensional momentum-space
setting~\cite{SSH1979,RiceMele1982}. For the SSH model, the elementary triangular construction is
replaced by a closed Bargmann product around the one-dimensional
Brillouin zone. Its accumulated phase captures the geometric phase associated with the
closed Bloch-state trajectory and, with the corresponding phase
unwrapping, reproduces the winding number of the SSH insulator. The Rice--Mele cycle then extends this construction to a
two-dimensional pumping parameter space, where the finite-state phase
accumulation yields $C_B=-1$. The resulting $\mathcal K_B$ resolves the
local distribution of the geometric phase across the parameter space,
while $\mathcal M_B$ characterizes the corresponding variation of the
finite-state overlap amplitudes, with both exhibiting pronounced
structure near the minimum-gap regions.

Taken together, the results show how finite-state quantum-state
overlaps can be used to extract both local geometric structure and
global Chern topology within a common framework. The third-order
Bargmann construction provides the elementary phase information for
the triangular finite-state geometry, while closed-loop and
composite-subspace extensions broaden the construction to
one-dimensional cycles and internally touching multiband systems.
Retaining both the phase and modulus of the overlap products further
provides complementary information on the finite-state geometry.
The resulting formulation connects discrete quantum-state data
directly to geometric and topological characteristics, without
requiring dynamical evolution or Berry connection or curvature as
inputs to the finite-state evaluation.
\section{Discrete Bargmann Geometry and Chern Topology}
\label{sec:bargmann_construction}

We formulate the finite-state construction by assigning gauge-invariant
geometric quantities to discrete quantum states sampled over a
two-dimensional parameter space. The elementary object is the
third-order Bargmann invariant, from which oriented geometric phases
can be associated with elementary triangles and subsequently combined
to characterize finite regions and their global topology.

For normalized states $|u_A\rangle$, $|u_B\rangle$, and $|u_C\rangle$
associated with three points in parameter space, the third-order
Bargmann invariant is given by \cite{Bargmann1964,Mukunda2003}
\begin{equation}
\Delta_3(A,B,C)
=
\langle u_A|u_B\rangle
\langle u_B|u_C\rangle
\langle u_C|u_A\rangle .
\label{eq:delta3_general}
\end{equation}
Under independent rephasing
$|u_j\rangle\rightarrow e^{i\chi_j}|u_j\rangle$, all phase factors
cancel, making $\Delta_3$ gauge invariant. For an individual Bloch band,
with $P_j=|u_j\rangle\langle u_j|$, the same quantity can be written as
$\Delta_3(A,B,C)
=
\operatorname{Tr}(P_AP_BP_C)$. This projector form makes explicit that the construction depends only
on the quantum states and their mutual overlaps, without requiring
derivatives of the Bloch states.

We consider a two-dimensional periodic system with an isolated
nondegenerate Bloch band,
\begin{equation}
H(\mathbf{k})|u_n(\mathbf{k})\rangle
=
E_n(\mathbf{k})|u_n(\mathbf{k})\rangle .
\label{eq:general_band}
\end{equation}
Its conventional Chern number is
\begin{equation}
C_n
=
\frac{1}{2\pi}
\int_{\mathrm{BZ}}d^2k\,\mathcal F_n(\mathbf{k}),
\label{eq:chern_definition}
\end{equation}
where $\mathcal F_n$ is the Berry curvature associated with the Berry
connection
$\mathcal A_\mu=i\langle u_n|\partial_{k_\mu}u_n\rangle$, with
$\mathcal F_n
=
\partial_{k_x}\mathcal A_y-\partial_{k_y}\mathcal A_x$
\cite{Berry1984,Thouless1982,Kohmoto1985}.
The same finite-state construction can be formulated more generally
for quantum states defined over a two-dimensional periodic parameter
space, subject to the corresponding spectral and finite-state
admissibility conditions.

To establish the finite-state correspondence, consider an oriented
triangle with vertices
$\mathbf{k}_A=\mathbf{k}$,
$\mathbf{k}_B=\mathbf{k}+\boldsymbol{\delta}_1$, and
$\mathbf{k}_C=\mathbf{k}+\boldsymbol{\delta}_2$, where
$\boldsymbol{\delta}_{1,2}=O(h)$ and $A_\triangle=O(h^2)$ denotes the
oriented area of the triangle. Expanding the three states about
$\mathbf{k}$, the gauge-dependent contributions cancel in the cyclic
product, while the leading gauge-invariant phase is determined by the
Berry curvature. For sufficiently small triangles,
\begin{equation}
\arg\Delta_3
=
-\mathcal F_n(\mathbf{k})A_\triangle
+O(h^3).
\end{equation}
We therefore define the oriented Bargmann phase as
\begin{equation}
\gamma_\triangle=-\arg\Delta_3 .
\label{eq:bargmann_triangle_phase}
\end{equation}
The minus sign here is a convention consistent with the orientation
and the Berry-connection convention adopted above; reversing the
orientation reverses the corresponding phase. With this convention,
\begin{equation}
\gamma_\triangle
=
\mathcal F_n(\mathbf{k})A_\triangle
+O(h^3).
\label{eq:bargmann_phase_flux}
\end{equation}
Thus, the third-order Bargmann phase provides a finite-state
representation of the Berry flux through an oriented elementary
triangle in the small-triangle limit.

\begin{figure}[H]
    \centering
    \includegraphics[width=0.52\linewidth]{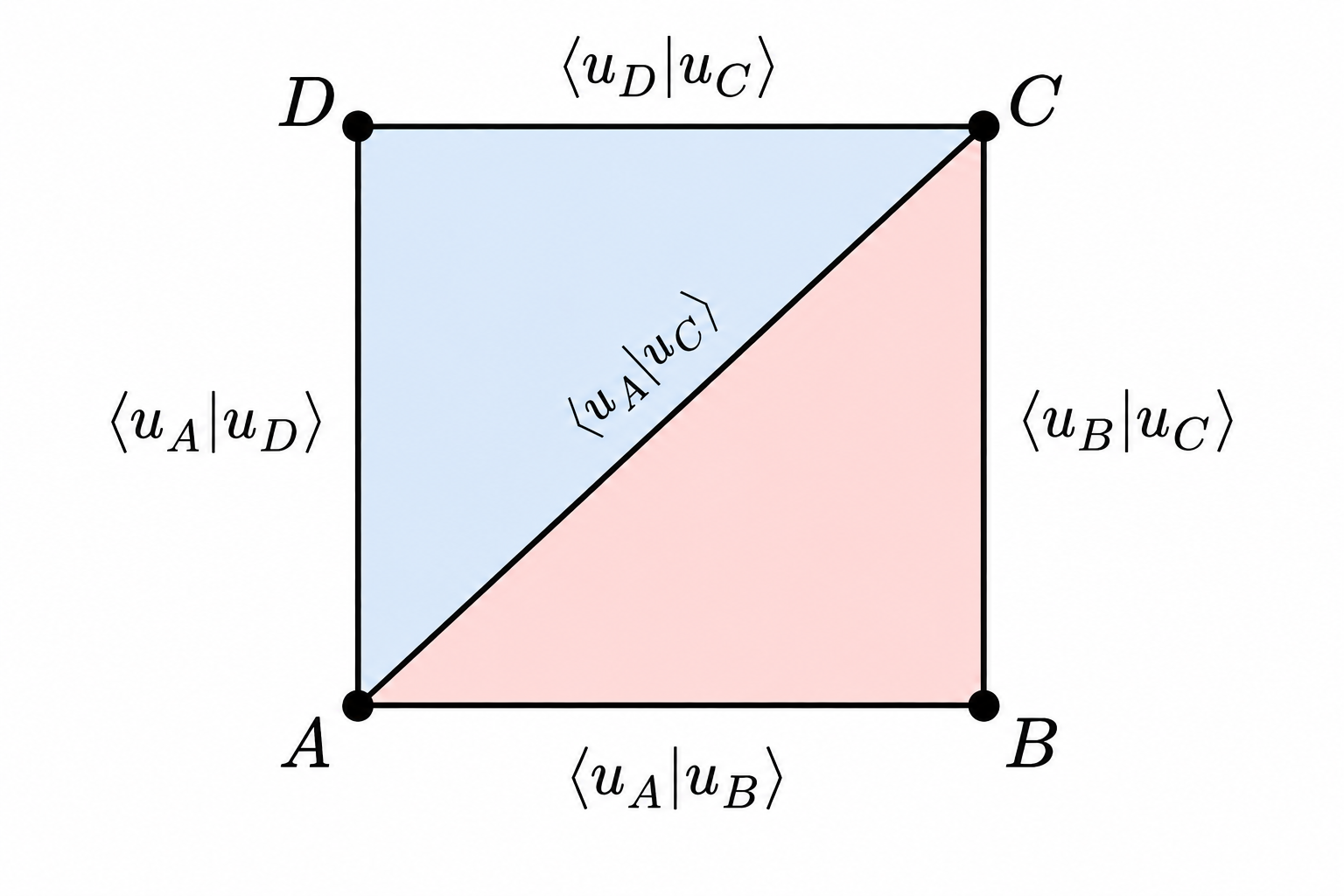}
    \caption{
        Finite Brillouin-zone cell constructed from the two oriented
        triangular third-order Bargmann invariants associated with
        $ABC$ and $ACD$. Each vertex represents the Bloch state at the
        corresponding Brillouin-zone point, while the links denote
        their mutual quantum-state overlaps. The two triangular
        contributions combine to give the geometric phase associated
        with the finite cell.
    }
    \label{fig:bargmann_cell}
\end{figure}

The construction can be applied directly to a triangulation of the
Brillouin zone. Let $A$, $B$, $C$, and $D$ denote neighboring
Brillouin-zone mesh points forming an oriented square cell, as
illustrated in Fig.~\ref{fig:bargmann_cell}, with corresponding Bloch
states $|u_A\rangle$, $|u_B\rangle$, $|u_C\rangle$, and $|u_D\rangle$.
Dividing the cell along the diagonal $AC$ gives the two consistently
oriented triangles $ABC$ and $ACD$. Since the overlap factors along the internal diagonal occur with
opposite orientations and combine as
$|\langle u_A|u_C\rangle|^2$, their phase contributions cancel, and
the corresponding cell phase satisfies
\begin{equation}
\gamma_{\square}
\equiv
\gamma_{ABC}+\gamma_{ACD}
\pmod{2\pi}.
\label{eq:square_phase_sum}
\end{equation}
For a sufficiently fine mesh, the phases can be represented on a
consistent branch connected to the small-cell limit, so that the
modulo-$2\pi$ ambiguity does not affect the local phase accumulation.
The elementary Bargmann phases are defined modulo $2\pi$, while the
accumulated phases used below are taken on a consistent continuous
branch.
The cell is thus constructed entirely from the same third-order
Bargmann invariant of Eq.~\eqref{eq:delta3_general}.

For a square cell of linear size $h$, with area $A_\square=h^2$, we
define the finite-resolution geometric density
\begin{equation}
\mathcal K_B(\mathbf{k};h)
=
\frac{\gamma_{\square}(\mathbf{k};h)}{A_\square}
=
-\frac{1}{h^2}
\arg\!\left[
\Delta_3^{ABC}\Delta_3^{ACD}
\right]
=
-\frac{1}{h^2}
\operatorname{Im}
\ln\!\left[
\Delta_3^{ABC}\Delta_3^{ACD}
\right].
\label{eq:finite_bargmann_curvature}
\end{equation}
Using Eq.~\eqref{eq:bargmann_phase_flux},
$\mathcal K_B(\mathbf{k};h)
=
\mathcal F_n(\mathbf{k})+O(h)$,
so that $\mathcal K_B$ provides a local finite-state geometric density
that approaches the Berry curvature in the fine-mesh limit. At finite
resolution, $\mathcal K_B$ is determined directly from the
quantum-state overlaps.

The same finite-state construction also retains amplitude information
through the modulus of the Bargmann invariants. We therefore define the
finite-resolution amplitude density
\begin{equation}
\mathcal M_B(\mathbf{k};h)
=
-\frac{1}{h^2}
\ln
\left|
\Delta_3^{ABC}\Delta_3^{ACD}
\right|^2 .
\label{eq:finite_bargmann_amplitude}
\end{equation}
The modulus is insensitive to the orientation of the triangles, so
$\mathcal M_B$ is independent of the orientation convention used for
$\mathcal K_B$ and satisfies $\mathcal M_B\geq0$ for normalized states.
For nearby states, the corresponding overlap expansion relates
$\mathcal M_B$ in the small-cell limit to the quantum metric, providing
a finite-state measure of the local quantum-state distance. Thus,
$\mathcal K_B$ and $\mathcal M_B$ retain complementary phase and
amplitude information about the local quantum-state geometry.

The global topology follows by accumulating the oriented Bargmann
phases over a closed periodic triangulation. For a sufficiently fine
admissible mesh, with the phases taken on a consistent branch, we define
the Bargmann-evaluated Chern number as
\begin{equation}
C_B=
\frac{1}{2\pi}
\sum_{\triangle}\gamma_\triangle .
\label{eq:CB_general}
\end{equation}
For a family of increasingly fine triangulations with characteristic
spacing $h$, Eq.~\eqref{eq:bargmann_phase_flux} gives
\begin{align}
\lim_{h\rightarrow0}
\frac{1}{2\pi}
\sum_{\triangle(h)}\gamma_\triangle
&=
\frac{1}{2\pi}
\lim_{h\rightarrow0}
\sum_{\triangle(h)}
\left[
\mathcal F_n(\mathbf{k}_\triangle)A_\triangle
+O(h^3)
\right]
\nonumber\\
&=
\frac{1}{2\pi}
\int_{\mathrm{BZ}}
d^2k\,\mathcal F_n(\mathbf{k})
=
C_n .
\label{eq:general_bargmann_chern_theorem}
\end{align}
Here the $O(h^3)$ term denotes the local remainder in the
small-triangle expansion. Since a fixed two-dimensional Brillouin zone contains
$O(h^{-2})$ triangles on a mesh of spacing $h$, its accumulated
contribution is $O(h)$ and therefore vanishes as $h\rightarrow0$.
The remaining area-weighted sum converges to the Berry-curvature
integral as a Riemann sum.

The Berry curvature therefore establishes the continuum correspondence
of the finite-state construction, while the finite-state evaluation
itself requires only gauge-invariant state overlaps. Equation
\eqref{eq:CB_general} consequently provides a finite-state
overlap-based evaluation of the conventional Chern number. The
individual cell phases retained in
Eq.~\eqref{eq:finite_bargmann_curvature} additionally preserve
finite-resolution local information in parameter space that is
compressed into the global Chern integer.

For an isolated nondegenerate band, the construction requires a
Bargmann-admissible triangulation,
$\Delta_3(A,B,C)\neq0$
for every elementary triangle. This condition guarantees that the
finite-state phase is well defined. It is distinct from spectral
isolation: a vanishing $\Delta_3$ concerns the chosen finite-overlap
construction and does not by itself imply a physical band-gap closing.
Conversely, the topological construction requires the relevant band to
remain isolated from the rest of the spectrum.

For a fixed finite mesh, $C_B$ is the phase sum defined in
Eq.~\eqref{eq:CB_general}. In general, its fine-mesh limit approaches
the conventional Chern number, as established in
Eq.~\eqref{eq:general_bargmann_chern_theorem}. For the QWZ and Hofstadter
examples below, the accumulated Bargmann phases yield the corresponding
integer Chern numbers already on the finite admissible meshes
considered. We therefore use $C_B$ throughout as the
Bargmann-evaluated Chern number.

The construction also extends to an isolated composite subspace. For
an $r$-dimensional subspace, let
$V_{\mathbf{k}}=(|u_1\rangle,\ldots,|u_r\rangle)$ be an orthonormal
local frame and define $M_{AB}=V_A^\dagger V_B$. The corresponding
finite-state invariant is
\begin{equation}
\Delta_3^{(r)}(A,B,C)
=
\det M_{AB}\det M_{BC}\det M_{CA}.
\label{eq:rank_r_bargmann}
\end{equation}
Under independent unitary changes of frame at the three vertices, the
determinant factors acquire compensating factors that cancel in the
closed product, so that $\Delta_3^{(r)}$ is invariant under the choice
of local frame for the composite subspace. For $r=1$, Eq.~\eqref{eq:rank_r_bargmann} reduces to
Eq.~\eqref{eq:delta3_general}. For an isolated composite subspace, the
corresponding phase sum approaches the first Chern number of the
subspace in the fine-mesh limit. This form is used below for the
central rank-$2$ subspace of the $1/4$-flux Hofstadter model, where the
individual central bands touch while their combined subspace remains
spectrally isolated.

Finally, the same overlap principle applies to an arbitrary closed
sequence of states,
\begin{equation}
B_N=
\prod_{j=1}^{N}\langle u_j|u_{j+1}\rangle,
\qquad
u_{N+1}=u_1 .
\label{eq:BN_general}
\end{equation}
When the overlaps are nonzero, we define the associated oriented
finite-loop phase by
$\gamma_N=-\arg B_N$,
with the same orientation convention as above, while its modulus
provides a separate finite-state amplitude quantity. This is the
$N$-state closed-loop generalization of the third-order Bargmann
construction, with $B_3=\Delta_3$. While the third-order Bargmann invariant captures the geometric information associated with an elementary three-state path, a general finite loop can retain information about the particular sequence of quantum states along the chosen closed path. This closed-loop form will be used below to examine
finite-path geometric information beyond the global Chern number.

The construction thus provides a common finite-state framework for
describing global Chern topology, local quantum geometry, and
closed-loop geometric information within the systems considered here. The following sections apply these constructions
to representative two- and multiband lattice models.
\section{Bargmann Geometry of the QWZ Chern Insulator}
\label{sec:qwz}

We first apply the finite-state Bargmann construction to the
Qi--Wu--Zhang (QWZ) model, which provides a minimal two-band
realization of a Chern insulator~\cite{QiWuZhang2006a,QiWuZhang2006b}.
It provides a simple setting in which the conventional Chern topology
can be compared directly with its finite-state Bargmann evaluation,
while the behavior of the geometric construction near topological
transitions can also be examined.

The Bloch Hamiltonian is
\begin{equation}
H(\mathbf{k})
=
\sin k_x\,\sigma_x
+
\sin k_y\,\sigma_y
+
\left(m+\cos k_x+\cos k_y\right)\sigma_z ,
\label{eq:qwz_hamiltonian}
\end{equation}
with $\mathbf{k}=(k_x,k_y)$ in the Brillouin zone
$-\pi\leq k_x,k_y<\pi$. It is convenient to define
$M(\mathbf{k})=m+\cos k_x+\cos k_y$,
so that the band energies are
$E_{\pm}(\mathbf{k})
=
\pm
\sqrt{
\sin^2k_x+\sin^2k_y+M^2(\mathbf{k})
}$.
The parameter $m$ controls the momentum-independent contribution to the
$\sigma_z$ component and therefore acts as a mass parameter for the
Dirac points. The lower band remains isolated throughout the Brillouin zone
provided $\mathbf d(\mathbf{k})\neq0$ for all $\mathbf{k}$, with
$\mathbf d=(\sin k_x,\sin k_y,M(\mathbf{k}))$.

The possible gap closings occur where the first two components of
$\mathbf d$ vanish, namely at the four high-symmetry points
$(0,0)$, $(\pi,0)$, $(0,\pi)$, and $(\pi,\pi)$. The corresponding
Dirac masses are $m+2$, $m$, $m$, $m-2$,
respectively, giving the critical values $m=-2,0,2$. For the
conventional continuum description, the Chern number is obtained by
computing the Berry curvature of the occupied band from the Bloch
eigenstates and integrating it over the Brillouin zone according to
Eq.~\eqref{eq:chern_definition}. Equivalently, the result can be
organized in terms of the contributions from the Dirac points, giving
for the lower band
\begin{equation}
C_-
=
\frac{1}{2}
\left[
\operatorname{sgn}(m+2)
-2\operatorname{sgn}(m)
+\operatorname{sgn}(m-2)
\right],
\label{eq:qwz_chern}
\end{equation}
and hence
\[
C_-=
\begin{cases}
0, & m<-2,\\
+1, & -2<m<0,\\
-1, & 0<m<2,\\
0, & m>2.
\end{cases}
\]
Thus, changes of topology occur when the bulk gap closes at
$m=-2,0,2$. The transition at $m=0$ involves the two symmetry-related
Dirac points $(\pi,0)$ and $(0,\pi)$ simultaneously, whereas the
transitions at $m=\pm2$ involve a single high-symmetry point. This
distinction will also be useful when examining the finite-state
Bargmann structure near the critical values.

We now evaluate the same topology from the quantum states on a
discrete Brillouin-zone triangulation. For three neighboring momenta,
the elementary object is the third-order Bargmann invariant introduced
in Eq.~\eqref{eq:delta3_general},
$\Delta_3
=
\langle u_-(\mathbf{k}_1)|u_-(\mathbf{k}_2)\rangle
\langle u_-(\mathbf{k}_2)|u_-(\mathbf{k}_3)\rangle
\langle u_-(\mathbf{k}_3)|u_-(\mathbf{k}_1)\rangle$.
Its gauge invariance makes it directly suitable for finite-state
evaluation: each triangle contributes the oriented phase
$\gamma_\triangle=-\arg\Delta_3$, as defined in
Eq.~\eqref{eq:bargmann_triangle_phase}. The global quantity is then
obtained from the phase sum of these elementary third-order
invariants according to Eq.~\eqref{eq:CB_general}.
Thus, the Chern topology of the QWZ band is obtained directly from the
phases of the finite-state overlap products, while the Berry curvature
provides the corresponding continuum reference.

As representative points in the gapped parameter regions, we perform a
numerical finite-state evaluation at $m=-1$ and $m=+1$, corresponding to
the two nontrivial Chern sectors, and at $m=-2.5$ and $m=+2.5$ in the
trivial regimes. For each value of $m$, the occupied-band eigenstates
are computed numerically on a finite admissible triangulation of the
Brillouin zone, and the third-order Bargmann invariant is evaluated on
each elementary triangle. The corresponding oriented phases are then
accumulated over the full Brillouin zone to obtain the finite-state
Chern number. The accumulated Bargmann phases are respectively
$2\pi$, $-2\pi$, $0$, and $0$, giving $C_B=+1$, $-1$, $0$, and $0$ for
these four representative values. Extending the same
finite-state calculation across the gapped parameter range gives
\begin{equation}
C_B=
\begin{cases}
0, & m<-2,\\
+1, & -2<m<0,\\
-1, & 0<m<2,\\
0, & m>2,
\end{cases}
\label{eq:qwz_CB_scan}
\end{equation}
on the finite admissible meshes considered here. The result follows
from the elementary third-order Bargmann phases themselves: the global
integer is obtained only after their oriented contributions are
accumulated over the Brillouin zone.

\begin{figure}[H]
    \centering
    \includegraphics[width=0.45\linewidth]{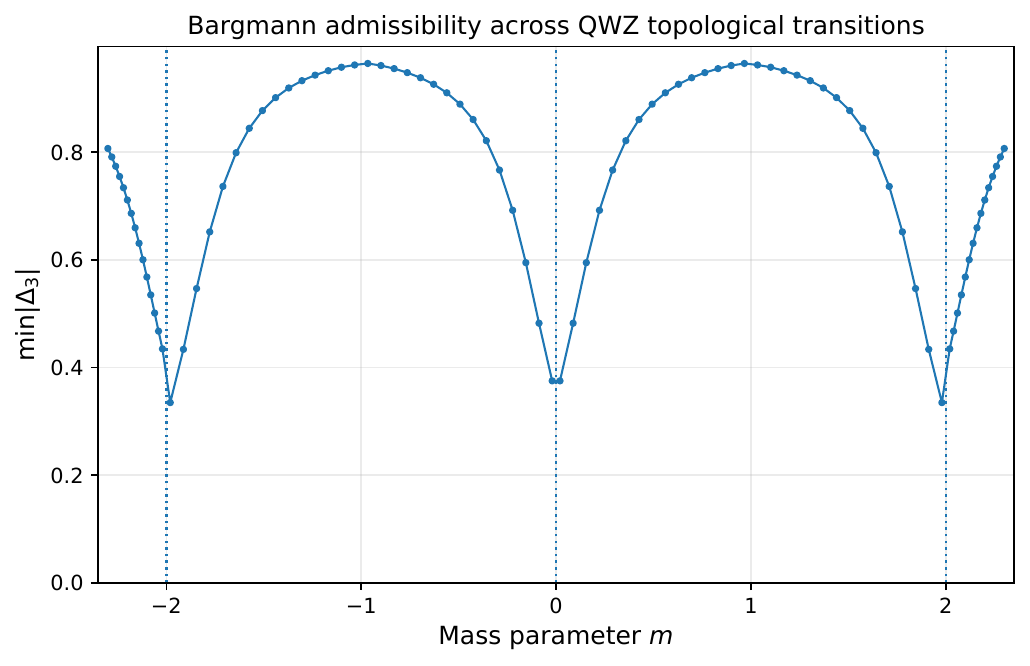}
    \caption{
Minimum magnitude of the elementary third-order Bargmann invariant,
$\Delta_{\min}(m)=\min_\triangle|\Delta_3|$, across the QWZ parameter
range. For the finite meshes considered, the suppression near the bulk
critical parameters accompanies the increasing variation of the Bloch
states as the band gap becomes small.
}
    \label{fig:qwz_bargmann_admissibility}
\end{figure}

The same elementary invariant also provides a useful measure of the
finite-state admissibility of the triangulation. We characterize it by
$\Delta_{\min}(m)
=
\min_{\triangle}|\Delta_3|$.
A nonzero value ensures that the phase of every elementary Bargmann
invariant is defined. For the meshes considered here, $\Delta_{\min}$
decreases as the system approaches $m=-2,0,2$, as shown in
Fig.~\ref{fig:qwz_bargmann_admissibility}. The decrease of $\Delta_{\min}$ near the critical parameters provides,
for this model and the meshes considered, a finite-state indication of
the changing overlap structure as the bulk gap-closing regions are
approached. The purpose of
$\Delta_{\min}$ here is to verify that the elementary Bargmann phases
remain well defined on the finite triangulation. Importantly,
$\Delta_3=0$ is a condition on the finite-state construction and does
not by itself constitute a criterion for spectral degeneracy. The three
critical values also need not produce identical minima in
$\Delta_{\min}$. At $m=0$, two symmetry-related Dirac points become
gapless simultaneously, whereas at $m=\pm2$ the closing occurs at a
single high-symmetry point. The corresponding finite-mesh overlap
structures therefore differ quantitatively near these transitions, as
reflected in the observed features of the $\Delta_{\min}(m)$ curve,
without implying a universal scaling law.

The finite-state construction can also be applied to closed paths in the
Brillouin zone. For the QWZ lower band, we evaluate the closed-loop
Bargmann product $B_N$ of Eq.~\eqref{eq:BN_general}, together with its
phase $\gamma_N=-\arg B_N$ and modulus $|B_N|$. The phase captures the
accumulated geometric phase along the finite sequence of states, while
the modulus characterizes the corresponding cumulative overlap
amplitudes. For circular and elliptical loops enclosing the same area,
we find, for the representative parameters considered,
$\gamma_N^{\mathrm{circ}}-\gamma_N^{\mathrm{ell}}
\simeq -1.84\times10^{-6}$, whereas
$|B_N|_{\mathrm{circ}}-|B_N|_{\mathrm{ell}}
\simeq 8.72\times10^{-4}$. Thus, although the accumulated phases are
nearly identical for these particular paths, their overlap moduli retain
a distinguishable finite-state dependence. This near equality of the
phases is specific to the chosen paths and parameter regime and does not
imply a general area-only dependence. More generally, the finite Bargmann loop retains information about the
sequence of quantum states along the path, with its phase approaching
the geometric flux through a sufficiently small loop, while its
modulus provides complementary information on the cumulative overlap
structure.

The QWZ model therefore establishes the finite-state Bargmann
evaluation of the conventional Chern topology while also illustrating
how the same overlap construction retains local finite-resolution
information. The suppression of
$\Delta_{\min}$ near the critical parameters characterizes the
admissibility of the finite-state triangulation, while the closed-loop
extension demonstrates that finite-state Bargmann quantities can also
retain information about the path followed by the quantum states.
\section{Bargmann Geometry of the $1/3$-Flux Hofstadter Model}
\label{sec:hofstadter13}

The QWZ model demonstrates the finite-state Bargmann construction in a
two-band system. We now apply the same framework to the $1/3$-flux
Hofstadter model\cite{Hofstadter1976}, providing a multiband setting in which global
topology and the local distribution of geometric structure can be
examined separately. The middle band remains spectrally isolated
throughout the magnetic Brillouin zone, so the rank-$1$ Bargmann
construction can be applied directly. At the same time, retaining the
individual finite-state phases before their global accumulation
provides a momentum-resolved geometric density whose local structure
is not contained in the final Chern integer.

For the flux considered here, the Bloch Hamiltonian is
\begin{equation}
H(k_x,k_y)=
\begin{pmatrix}
-2\cos\!\left(k_y-\frac{2\pi}{3}\right) & -1 &
-e^{-3ik_x}\\
-1 & -2\cos\!\left(k_y-\frac{4\pi}{3}\right) & -1\\
-e^{3ik_x} & -1 & -2\cos k_y
\end{pmatrix},
\label{eq:hofstadter13_H}
\end{equation}
with magnetic Brillouin zone
$0\leq k_x<2\pi/3$ and $0\leq k_y<2\pi$. The three band energies are
determined by
$E^3-6E+2\cos(3k_x)+2\cos(3k_y)=0$.
The middle band, $n=2$, is isolated throughout the magnetic Brillouin
zone.

We discretize the magnetic Brillouin zone by a uniform square mesh
with
$h=\Delta k_x=\Delta k_y=\frac{2\pi}{3N_B}$,
using $N_B$ divisions along $k_x$ and $3N_B$ along $k_y$, with
periodic boundary conditions. Each square is divided into two
counter-clockwise oriented triangles. The normalized middle-band
eigenstates at the mesh vertices are then used in the third-order
Bargmann construction of Sec.~\ref{sec:bargmann_construction}. The
oriented triangle phases and their global accumulation are evaluated
according to Eqs.~\eqref{eq:bargmann_triangle_phase} and
\eqref{eq:CB_general}, respectively.
For every admissible mesh considered here, the accumulated Bargmann
phase is
$\sum_{\triangle}\gamma_\triangle=-4\pi$,
giving
$C_B=-2$.

The same value $C_B=-2$ is obtained for all admissible meshes
examined, including $N_B=3,4,6,9,12,18,$ and $27$. The coarsest tested case, $N_B=2$, is excluded because one elementary
invariant becomes numerically indistinguishable from zero,
$\min|\Delta_3|\simeq3.1\times10^{-17}$, making its phase numerically
unreliable. This distinguishes the spectral isolation of the
middle band from the admissibility of a particular finite-state
triangulation.

The global phase sum, however, compresses the cell-resolved geometric
information into a single integer. We therefore evaluate the
finite-resolution geometric density defined in
Sec.~\ref{sec:bargmann_construction} from the two oriented triangular
phases associated with each square cell. For the present uniform mesh,
the cell area is $h^2$, so that $\mathcal K_B$ is evaluated directly
from the neighboring middle-band eigenstates on each cell.

\begin{figure*}[!t]
    \centering

    \begin{minipage}{0.48\textwidth}
        \centering
        \includegraphics[width=\linewidth]
        {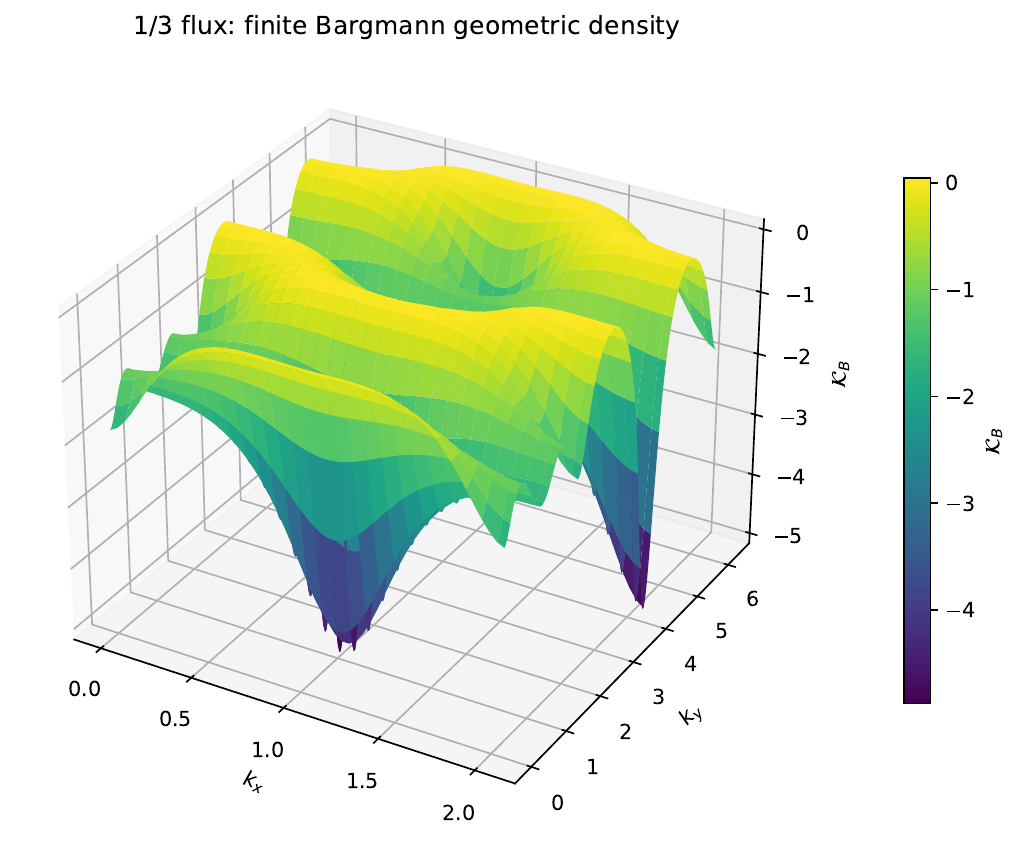}
        
        \vspace{2mm}
        \centering
        \text{(A)}
    \end{minipage}
    \hfill
    \begin{minipage}{0.48\textwidth}
        \centering
        \includegraphics[width=\linewidth]
        {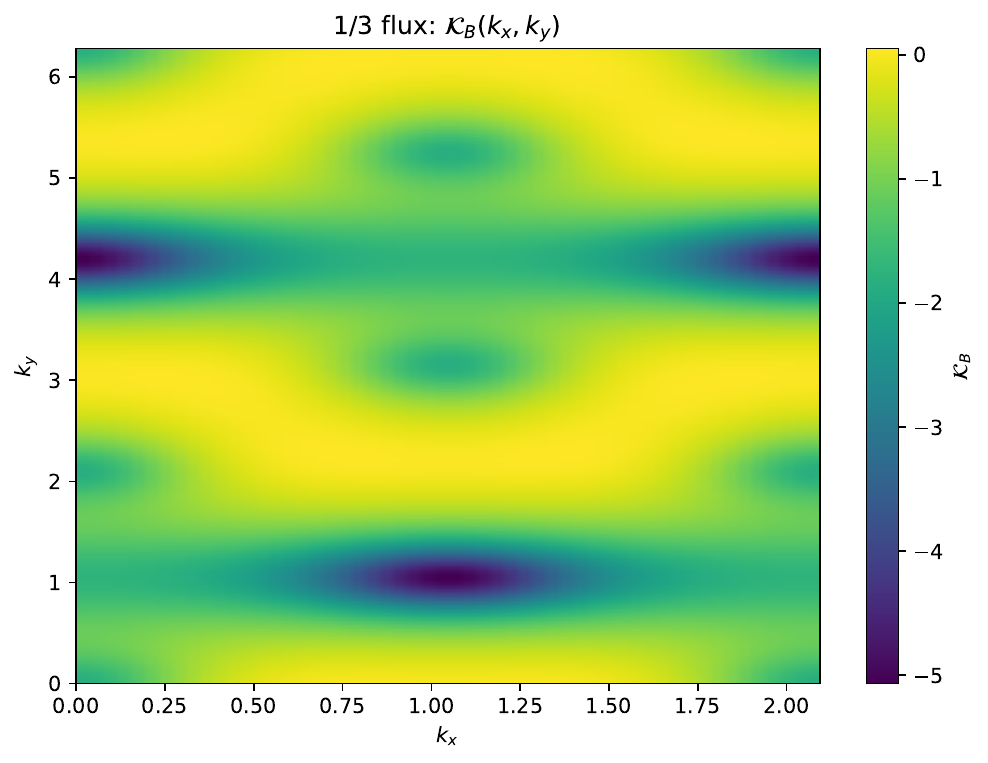}
        
        \vspace{2mm}
        \centering
        \text{(B)}
    \end{minipage}

    \vspace{5mm}

    \begin{minipage}{0.48\textwidth}
        \centering
        \includegraphics[width=\linewidth]
        {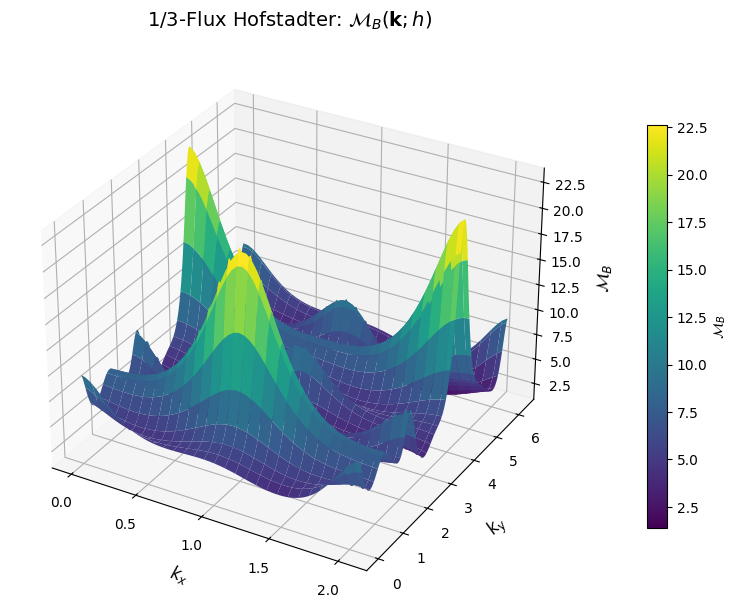}
        
        \vspace{2mm}
        \centering
        \text{(C)}
    \end{minipage}
    \hfill
    \begin{minipage}{0.48\textwidth}
        \centering
        \includegraphics[width=\linewidth]
        {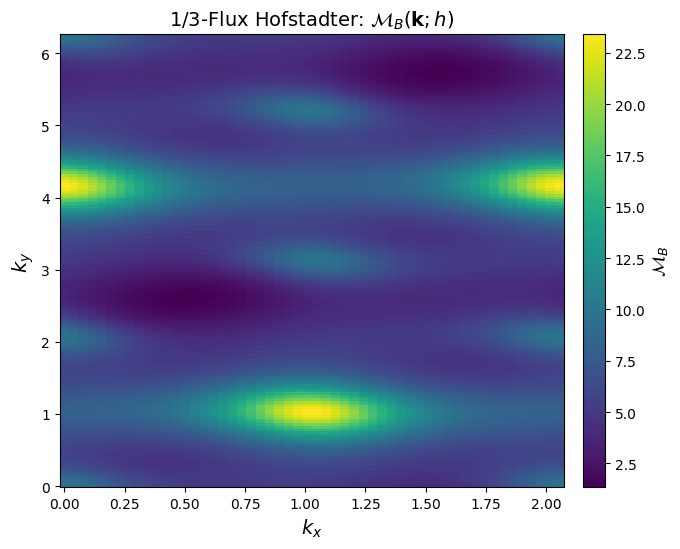}
        
        \vspace{2mm}
        \centering
        \text{(D)}
    \end{minipage}

    \caption{
    Finite-state Bargmann geometry of the middle band in the
    $1/3$-flux Hofstadter model for $N_B=54$. Panel (A) shows the
    three-dimensional cell-resolved Bargmann geometric density
    $\mathcal K_B(\mathbf{k};h)$, while panel (B) shows its distribution
    over the magnetic Brillouin zone. Panel (C) presents the
    three-dimensional cell-resolved amplitude density
    $\mathcal M_B(\mathbf{k};h)$, and panel (D) shows its corresponding
    distribution over the magnetic Brillouin zone. The $\mathcal K_B$
    panels display the phase sector of the finite-state geometry,
    whereas the $\mathcal M_B$ panels display the complementary
    amplitude sector obtained from the modulus of the same Bargmann
    construction. The pronounced extrema in the two sectors identify
    regions of strong finite-state geometric variation, while the
    corresponding momentum-space distributions retain information that
    is not contained in the global Chern integer.
    }
    \label{fig:hof13_local_geometry}
\end{figure*}

For $N_B=54$, the resulting $\mathcal K_B$ is strongly nonuniform
across the magnetic Brillouin zone, with
$-5.07\lesssim\mathcal K_B\lesssim0.053$.
The corresponding heat map and three-dimensional surface are shown
together with the complementary amplitude-sector density in
Fig.~\ref{fig:hof13_local_geometry}. The pronounced negative extrema
identify specific regions of the magnetic Brillouin zone where the
finite-resolution geometric density is particularly large in magnitude.
Their detailed momentum-space distribution is lost when the cell
phases are accumulated into the single integer $C_B=-2$.

The total phase also fixes the Brillouin-zone average of this local
density. Since the magnetic Brillouin-zone area is
$A_{\mathrm{MBZ}}=4\pi^2/3$,
\begin{equation}
\left\langle\mathcal K_B\right\rangle
=
\frac{2\pi C_B}{A_{\mathrm{MBZ}}}
=
-\frac{3}{\pi}.
\label{eq:hof13_KB_average}
\end{equation}
For $N_B=54$, the numerical average is
$-0.954929658551$, in agreement with $-3/\pi$ to numerical precision.
Thus, the local Bargmann density preserves the global phase constraint
while resolving how the geometric contribution is distributed across
momentum space.

The extrema of this finite-state map also provide useful locations for
a more detailed microscopic analysis. One representative minimum
occurs at
$\mathbf{k}_0=(0,4\pi/3)$.
At this point,
\[
H(\mathbf{k}_0)=
\begin{pmatrix}
1&-1&-1\\
-1&-2&-1\\
-1&-1&1
\end{pmatrix},
\]
with energies $E_1=-1-\sqrt3$,
$E_2=\sqrt3-1$,
$E_3=2$.
Both neighboring gaps are finite,
$E_2-E_1=2\sqrt3$,
$E_3-E_2=3-\sqrt3$,
so the prominent local structure is not associated with a band
degeneracy at this point or with a loss of spectral isolation of the
middle band. A symmetry-related extremal point occurs at
$\mathbf{k}_1=(\pi/3,\pi/3)$.

Having localized these regions directly from the finite-state Bargmann
density, we can then examine their microscopic origin using the
continuum multiband geometric response. Evaluating the standard multiband curvature expression for the middle
band at the representative point $\mathbf{k}_0$, the contribution from
the lower band vanishes, while the coupling to the upper band gives
$\mathcal F_2(\mathbf{k}_0)
=
-\frac{5+3\sqrt3}{2}$.
The symmetry-related point $\mathbf{k}_1$ gives the same extremal
curvature. Thus, the pronounced finite-state feature identifies a
region of strong local geometric response even in the absence of a
local
band degeneracy, while the continuum expression resolves this response
in terms of the interband couplings and energy separations.

This ordering also provides a useful way to organize the analysis of
multiband systems. The finite-state Bargmann density can first provide
a momentum-space map of prominent geometric regions using only
discrete eigenstate overlaps. A continuum analysis can then be applied
selectively at representative extrema or symmetry-related points to
analyze the microscopic origin of the observed structure. In this
sense, the Bargmann construction provides a finite-state localization
of geometric structure, while the continuum description supplies its
differential interpretation. The two descriptions are therefore
complementary rather than interchangeable.

The same finite-state construction also retains amplitude information
through the modulus of the Bargmann invariants. The corresponding
finite-resolution amplitude density $\mathcal M_B$ is defined in
Sec.~\ref{sec:bargmann_construction}. It characterizes the local
variation of the finite-state overlap amplitudes. Thus, while
$\mathcal K_B$ characterizes the phase sector of the finite-state
geometry, $\mathcal M_B$ characterizes its complementary amplitude
sector.
For the same $N_B=54$ mesh, $\mathcal M_B$ is also strongly nonuniform
across the magnetic Brillouin zone, with
$1.3333\lesssim\mathcal M_B\lesssim23.3998$,
and numerical average
$\langle\mathcal M_B\rangle=6.57495$.
The corresponding heat map and three-dimensional surface are shown in
Fig.~\ref{fig:hof13_local_geometry}(C,D). The pronounced maxima
identify regions where the finite-state overlap amplitudes are most
strongly reduced, indicating stronger variation of the middle-band
quantum states across the finite cells. These regions occur near
prominent structures of the $\mathcal K_B$ distribution, reflecting
their common origin in the same finite-state quantum geometry, while
retaining the distinct amplitude information contained in the modulus
of the Bargmann invariant.

To further characterize the amplitude-sector structure, we examine the
mesh cell for which the maximum value
$\mathcal M_B=23.3998$ is obtained, at
$(k_x,k_y)=(1.0472,1.0084)$ on the $N_B=54$ mesh. The middle-band
energies at this point are $E_1=-2.0023$, $E_2=-0.7290$, and
$E_3=2.7312$, giving gaps of $1.2733$ and $3.4602$ to the adjacent
bands. Thus, at this location, the large $\mathcal M_B$ value is not
accompanied by an anomalously small spectral gap. Instead, it reflects
the cumulative variation of the middle-band eigenstates across the
finite cell, with the two triangular contributions giving $11.6750$
and $11.7249$, respectively. The location lies close to the
symmetry-related region of pronounced $\mathcal K_B$ structure around
$(\pi/3,\pi/3)$, indicating that the phase and amplitude sectors can
both exhibit pronounced structure in regions of strong local
state-space variation.

The $1/3$-flux Hofstadter model therefore provides a multiband
realization in which the global Bargmann phase sum and the
cell-resolved geometric and amplitude densities can be examined within
the same finite-state construction. The middle-band result $C_B=-2$
is obtained on the admissible finite meshes considered, while the
resolved densities reveal finite-resolution structure across the
magnetic Brillouin zone that can be further interpreted using the
multiband Hamiltonian. We next consider the
$1/4$-flux model, where the individual central bands touch and the
composite-subspace extension of the Bargmann construction becomes
necessary.
\section{Bargmann Geometry of the $1/4$-Flux Hofstadter Model}
\label{sec:hofstadter14}

The $1/4$-flux Hofstadter model extends the finite-state construction
to a setting in which an individual band is not globally isolated.
The two central bands touch at isolated momenta, while the central pair
remains separated from the outer bands. This provides two distinct but
complementary applications of the Bargmann construction. Locally, the
rank-one invariant can be examined in a punctured neighborhood of the
band touching, where the individual eigenstates remain well defined.
Globally, however, the appropriate object is the isolated
two-dimensional central subspace. The model therefore connects the
local rank-one state geometry near an internal degeneracy with the
global finite-state topology of the corresponding composite subspace.

For flux $1/4$, the Bloch Hamiltonian is
\begin{equation}
H(\mathbf{k})=
\begin{pmatrix}
-2\cos k_y & -1 & 0 & -e^{-4ik_x}\\
-1 & -2\sin k_y & -1 & 0\\
0 & -1 & 2\cos k_y & -1\\
-e^{4ik_x} & 0 & -1 & 2\sin k_y
\end{pmatrix},
\label{eq:hof14_H}
\end{equation}
with magnetic Brillouin zone
$0\leq k_x<\pi/2$ and $0\leq k_y<2\pi$. Its spectrum satisfies
$E^4-8E^2+4-2\cos(4k_x)-2\cos(4k_y)=0$.
At the representative point $\mathbf{k}_*=(0,0)$, the two central
bands touch, with
$E_1=-2\sqrt2$, $E_2=E_3=0$, and $E_4=2\sqrt2$, while the central pair
remains separated from the outer bands throughout the magnetic
Brillouin zone. Consequently, the individual central bands cannot be
regarded as globally isolated rank-one bands, whereas their combined
two-dimensional subspace remains globally isolated.

To examine the local state geometry near the touching, consider the
representative point $\mathbf{k}_*=(0,0)$. Projecting the Hamiltonian
near $\mathbf{k}_*$ onto the central two-state sector, in a suitable
basis of the degenerate subspace, gives
$h_{\mathrm{eff}}(\mathbf q)
=
\sqrt2\,q_x\sigma_y
+\frac{\sqrt2}{3}q_y\sigma_x
-\frac{4}{3}q_y\sigma_z$,
and hence
$E_\pm(\mathbf q)=\pm\sqrt2\,|\mathbf q|+O(q^2)$.
The central bands therefore separate linearly away from the touching.

\begin{figure}[H]
    \centering
    \begin{minipage}[t]{0.45\linewidth}
        \centering
        \includegraphics[width=\linewidth]
        {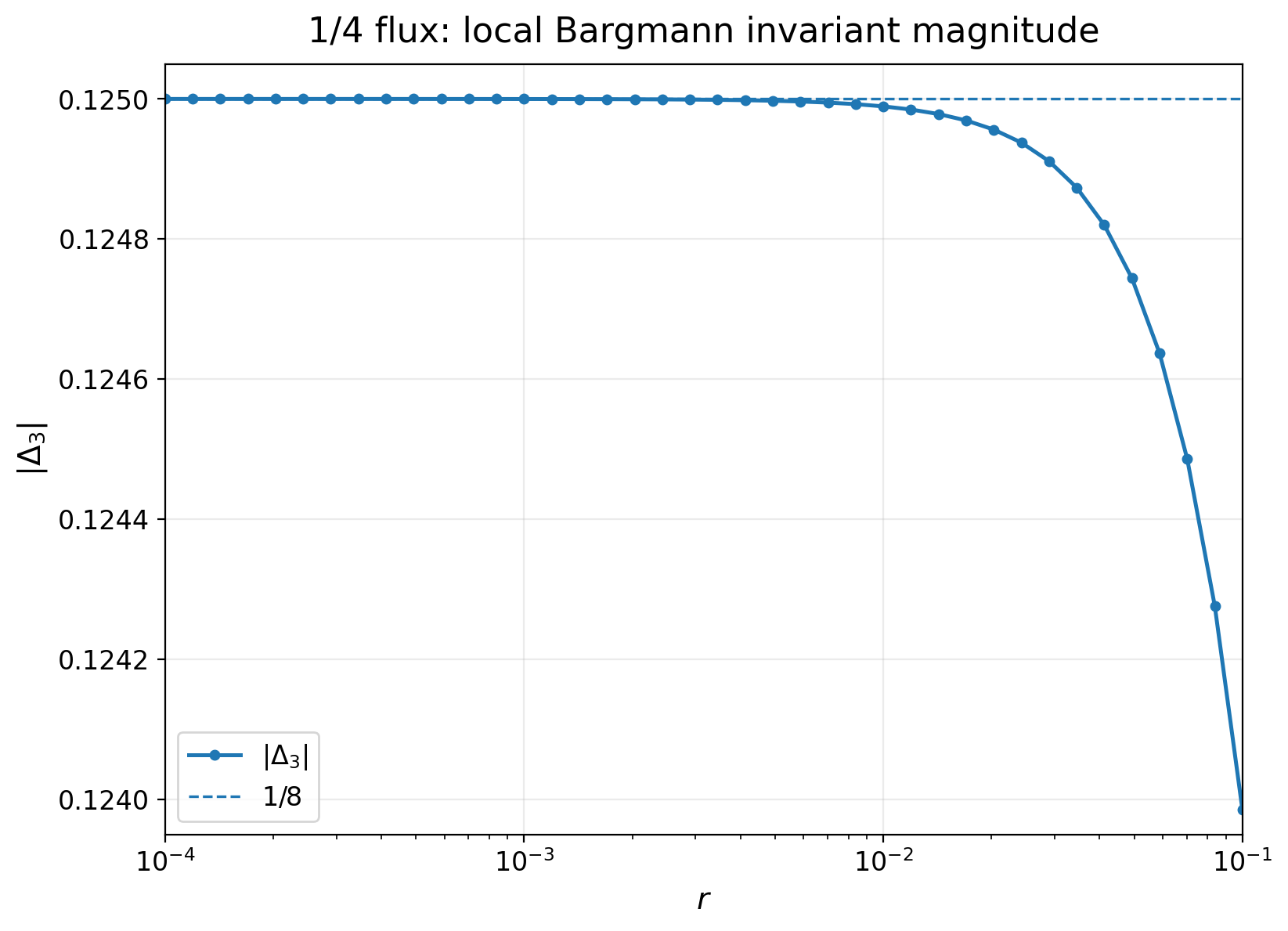}

        \vspace{1mm}
        \text{(A)}
    \end{minipage}
    \hfill
    \begin{minipage}[t]{0.44\linewidth}
        \centering
        \includegraphics[width=\linewidth]
        {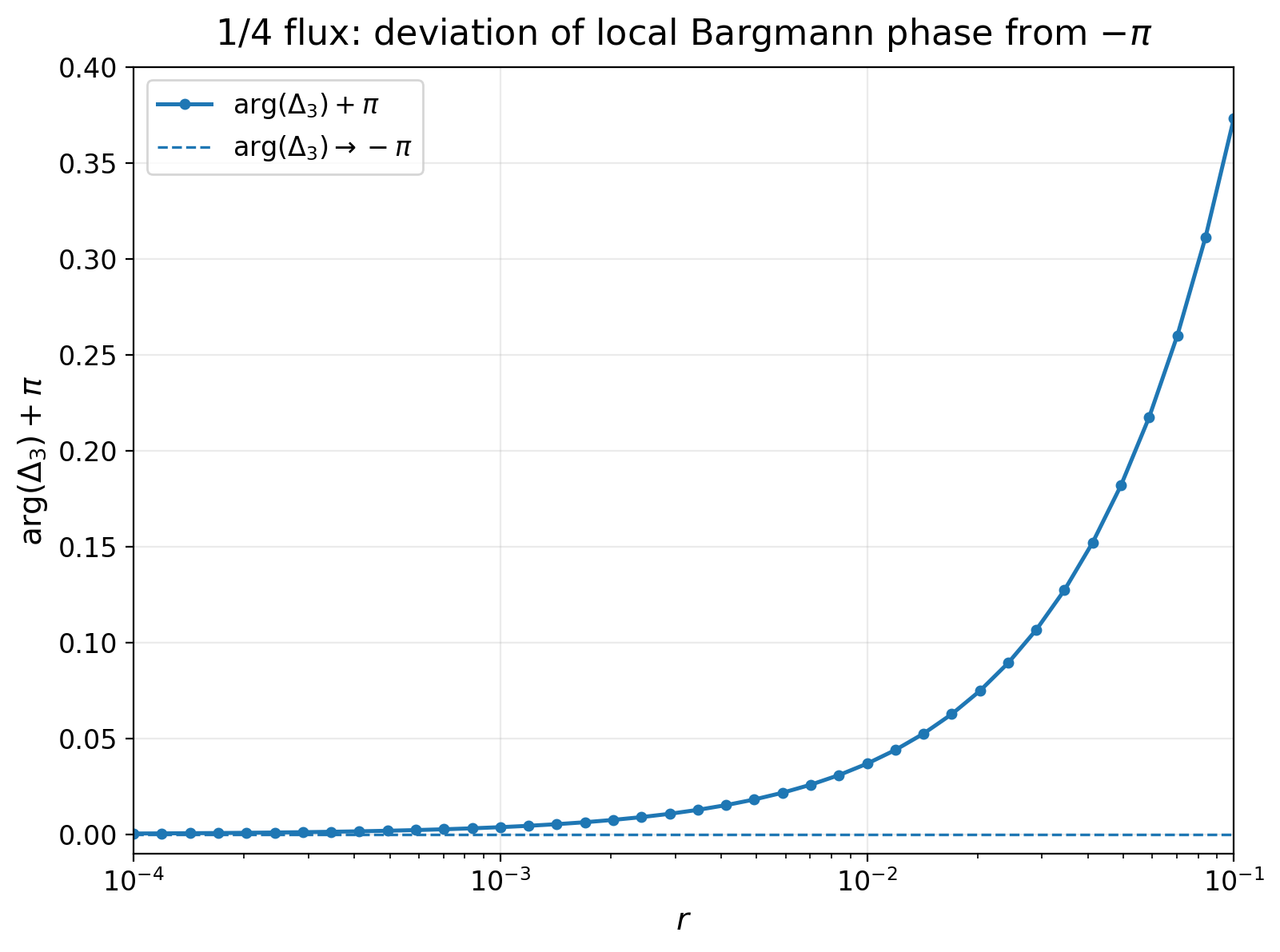}

        \vspace{1mm}
        \text{(B)}
    \end{minipage}

    \caption{
    Local rank-one Bargmann characterization of the central-band
    touching in the $1/4$-flux Hofstadter model. Panel (A) shows the
    magnitude of the third-order Bargmann invariant for the three-state
    construction on a small loop of radius $r$ around
    $\mathbf{k}_*=(0,0)$; as $r$ decreases, $|\Delta_3|$ approaches the
    analytical limiting value $1/8$. Panel (B) shows the corresponding
finite-radius deviation of the Bargmann phase from its limiting value
$-\pi$ on the chosen continuous branch, plotted as
$\arg(\Delta_3)+\pi$; the quantity approaches zero as
$r\rightarrow0$, consistent with
$\arg\Delta_3\rightarrow-\pi$ on this branch and the limiting value
$\Delta_3=-1/8$. Together, the two panels show the magnitude and
    phase behavior of the same local rank-one Bargmann construction
    near the internal central-band touching.
    }
    \label{fig:hof14_local_delta3}
\end{figure}

This local linear structure produces a characteristic finite-state
signature in the third-order Bargmann invariant. Consider a small
circle of radius $r$ around $\mathbf{k}_*$ and choose three states at
equal angular intervals along the resulting effective Bloch circle.
Their normalized Bloch vectors lie on a fixed great circle, with
pairwise scalar products $-1/2$ and vanishing scalar triple product.
Using
$\Delta_3
=
\frac14\left[
1+\mathbf n_A\!\cdot\mathbf n_B
+\mathbf n_B\!\cdot\mathbf n_C
+\mathbf n_C\!\cdot\mathbf n_A
+i\,\mathbf n_A\!\cdot
(\mathbf n_B\times\mathbf n_C)
\right]$,
one obtains
$\Delta_3\rightarrow-1/8$ in the $r\rightarrow0$ limit, with
$\arg\Delta_3\rightarrow\pi\equiv-\pi\pmod{2\pi}$.
For a finite-radius loop with the same angular prescription, the full
four-band calculation approaches this limit as
$\Delta_3(r)=-1/8+O(r)$.

The corresponding finite-radius behavior is shown in
Fig.~\ref{fig:hof14_local_delta3}. The magnitude approaches the nonzero limiting value $1/8$ as the loop
shrinks, while the phase approaches $-\pi$ on the continuous branch
chosen for the finite-radius calculation. To make the small
finite-radius variation of the phase explicit, panel (B) plots
$\arg(\Delta_3)+\pi$, which approaches zero as $r\rightarrow0$ on this
branch. The limiting value $\Delta_3=-1/8$ lies on the branch boundary
$\pi\equiv-\pi$ of the phase. The rank-one finite-state construction
therefore remains well defined for nonzero distance from the
degeneracy, even though the individual central-band eigenstate becomes
non-unique exactly at the touching. This provides a concrete local
Bargmann signature associated with the chosen loop around the internal
band touching, while the result by itself does not imply that a
vanishing or nonvanishing $\Delta_3$ constitutes a general criterion
for band degeneracy.

The local rank-one description, however, cannot be continued through
the degeneracy as a globally defined individual-band construction.
The appropriate global object is instead the isolated central
two-dimensional subspace, with projector
$P_{23}(\mathbf{k})
=
|u_2(\mathbf{k})\rangle\langle u_2(\mathbf{k})|
+
|u_3(\mathbf{k})\rangle\langle u_3(\mathbf{k})|$,
where, at the touching, the two vectors may be replaced by any
orthonormal basis spanning the same central subspace. At each momentum,
we choose an arbitrary orthonormal frame
$V(\mathbf{k})
=
\bigl(|v_1(\mathbf{k})\rangle,|v_2(\mathbf{k})\rangle\bigr)$
spanning this subspace. In particular, the frame is not required to
coincide with the individual eigenstates at the internal touching.

For two momenta $A$ and $B$, define the overlap matrix
$M_{AB}=V_A^\dagger V_B$. The rank-$2$ Bargmann invariant is the
specialization of the composite-subspace construction introduced in
Sec.~\ref{sec:bargmann_construction} as,
$\Delta_3^{(2)}(A,B,C)
=
\det M_{AB}\,
\det M_{BC}\,
\det M_{CA}$.
The resulting closed determinant product is invariant under independent
unitary changes of frame at the three momenta and therefore depends only
on the central composite subspace, not on the particular choice of local
basis used to represent it.

The rank-$2$ invariant is incorporated into the same oriented
phase-summation construction introduced in
Sec.~\ref{sec:bargmann_construction}. Thus, the global topology of the
central pair is evaluated from the phases of the rank-$2$ Bargmann
invariants accumulated over the triangulated magnetic Brillouin zone,
with the composite-subspace invariant replacing the rank-one
invariant. The overall sign follows the orientation and phase
convention used throughout the finite-state calculation.

For the meshes
$(N_x,N_y)=(3,12),(4,16),(6,24)$, and $(9,36)$, all elementary
rank-$2$ invariants remain nonzero. In each case, the oriented phases
accumulate to
\begin{equation}
\sum_{\triangle}\gamma^{(2)}_\triangle=-4\pi,
\qquad
C_{23}=-2 .
\label{eq:hof14_C23_result}
\end{equation}
The same integer is therefore obtained on all finite meshes considered.
The minimum magnitude of the rank-$2$ invariant increases from
approximately $0.1603$ on the coarsest mesh to $0.8202$ on the finest,
so that the elementary rank-$2$ phases remain well defined on every
mesh used for the global calculation.

\begin{figure}[H]
    \centering
    \includegraphics[width=0.48\linewidth]
    {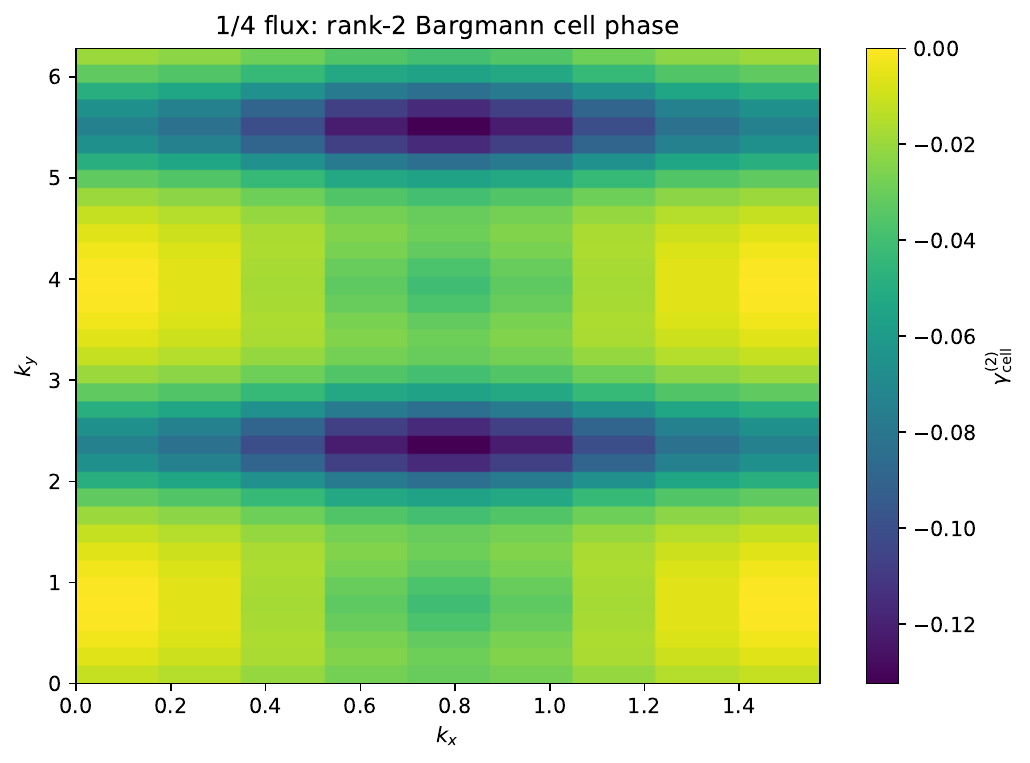}
    \caption{
    Local rank-$2$ Bargmann phase of the central composite subspace in
    the $1/4$-flux Hofstadter model over the magnetic Brillouin zone.
    The rank-$2$ invariants remain nonzero on the mesh considered, so
    that the finite-cell phases are well defined, and their oriented
    accumulation gives $C_{23}=-2$.
    }
    \label{fig:hof14_rank2_phase}
\end{figure}
The corresponding local rank-$2$ Bargmann phase distribution is shown
in Fig.~\ref{fig:hof14_rank2_phase}. For the $9\times36$ mesh used
here, the cell-resolved phase ranges approximately from
$\gamma_{\mathrm{cell}}^{(2)}=-0.132$ to $0$. One extremum is visible
near $(k_x,k_y)\simeq(0.785,2.269)$, together with a
symmetry-related feature near $(k_x,k_y)\simeq(0.785,5.411)$.
These features indicate regions in momentum space where the isolated
central composite subspace accumulates larger finite-state geometric
phases. Their symmetry-related appearance reflects the corresponding
symmetry of the composite-state geometry, while their oriented
accumulation over the full magnetic Brillouin zone gives the global
Chern number $C_{23}=-2$. Thus, the rank-$2$ phase map provides a
finite-resolution view of the local distribution of the geometric
phase of the composite subspace, even though the individual central
bands are not globally isolated.

The $1/4$-flux Hofstadter model therefore demonstrates how the
finite-state construction adapts when an internal band touching
obstructs a global rank-one description. The specified
punctured-neighborhood construction gives the local limiting behavior
$\Delta_3\rightarrow-1/8$, while the isolated central subspace provides
the appropriate global object and yields $C_{23}=-2$ through the
rank-$2$ construction. The local rank-one and global composite-subspace
descriptions thus probe complementary aspects of the same finite-state
geometry.
\section{Finite-State Bargmann Geometry in SSH and Rice--Mele Models}
\label{sec:ssh_rice_mele}

The finite-state construction developed in
Sec.~\ref{sec:bargmann_construction} also applies to one-dimensional
cycles and to parameter spaces generated by adiabatic pumping. The
Su--Schrieffer--Heeger (SSH) model provides the one-dimensional
setting, where the relevant finite-state object is a closed product of
successive state overlaps. Introducing a cyclic modulation of the
dimerization together with a staggered potential gives the Rice--Mele
pump, in which momentum and the pumping parameter form a
two-dimensional torus. The same third-order Bargmann construction can
then be applied to this parameter space, yielding both the global
topological invariant and a cell-resolved finite-state geometric
density.

\subsection{SSH Model: Closed Bargmann Loops and Winding}
\label{subsec:ssh}

We first consider the standard Su--Schrieffer--Heeger (SSH) model
\cite{SSH1979} as a one-dimensional realization of the finite-state
overlap construction. Unlike the two-dimensional Brillouin-zone
construction, where the elementary geometric object is associated with
a triangle, the natural finite-state object in one dimension is the
closed product of overlaps along the Brillouin-zone cycle. This
provides a finite-state characterization of the winding
information carried by the occupied Bloch state.

The SSH Hamiltonian is
\begin{equation}
H_{\mathrm{SSH}}(k)=
\begin{pmatrix}
0 & t_1+t_2e^{-ik}\\
t_1+t_2e^{ik} & 0
\end{pmatrix}
=
\mathbf d(k)\cdot\boldsymbol{\sigma},
\label{eq:ssh_hamiltonian}
\end{equation}
with
$\mathbf d(k)=
\left(
t_1+t_2\cos k,\,
t_2\sin k,\,
0
\right)$. The band energies are
$E_{\pm}(k)
=
\pm\sqrt{t_1^2+t_2^2+2t_1t_2\cos k}$,
so that, for positive hoppings, the bulk gap closes at
$t_1=t_2$ and $k=\pi$. The two gapped regimes are distinguished by
the winding of $\mathbf d(k)$ around the origin.

For the occupied band, we choose
$|u_-(k)\rangle=
\frac{1}{\sqrt{2}}
\begin{pmatrix}
-e^{-i\phi(k)}\\
1
\end{pmatrix}$,
$\phi(k)=\arg\!\left(t_1+t_2e^{ik}\right)$.
The corresponding Bloch vector lies on the equator of the Bloch sphere.
For neighboring states, the elementary third-order Bargmann phase does
not by itself capture the winding accumulated over the full
Brillouin-zone cycle.
Instead, that information is retained by the closed overlap product
around the complete cycle.

Following the general closed-loop construction of
Eq.~\eqref{eq:BN_general}, we define
\[
B_N^{\mathrm{SSH}}
=
\prod_{j=1}^{N}
\langle u_-(k_j)|u_-(k_{j+1})\rangle,
\]
with $k_{N+1}=k_1$.
The product is invariant under independent rephasing of the states at
the mesh points. Its phase gives the accumulated finite-state
geometric phase around the Brillouin zone, while its modulus provides
the corresponding finite-overlap magnitude. With the orientation
convention of Sec.~\ref{sec:bargmann_construction}, we define
$\gamma_{\mathrm{BZ}}^{\mathrm{SSH}}
=
-\arg B_N^{\mathrm{SSH}}$.
For the occupied states above, the accumulated phase of the closed
product approaches one half of the unwrapped change of $\phi(k)$ around
the Brillouin zone. In the fine-mesh limit,
$\gamma_{\mathrm{BZ}}^{\mathrm{SSH}}
=
\frac{1}{2}\Delta_{\mathrm{wind}}\phi
=
\pi W$, where
$\Delta_{\mathrm{wind}}\phi=2\pi W$
is the unwrapped change of $\phi(k)$ over one traversal of the
Brillouin zone. Thus,
$W=\frac{\gamma_{\mathrm{BZ}}^{\mathrm{SSH}}}{\pi}$,
giving $W=0$ for $t_1>t_2$ and $W=1$ for $t_2>t_1$ for positive
hoppings. Here the use of the unwrapped phase change is essential:
the winding is determined by the total continuous phase accumulation
along the Brillouin-zone cycle, rather than by the difference of
principal values of $\phi(k)$ at the endpoints.

The SSH model therefore provides the one-dimensional counterpart of
the finite-state construction: the closed overlap product follows the
complete Brillouin-zone cycle, with its phase retaining the winding
information of the occupied Bloch state. This one-dimensional
construction complements the triangular Bargmann formulation used for
two-dimensional parameter spaces.

\subsection{Rice--Mele Model: Finite-State Chern Geometry}
\label{subsec:rice_mele}

A two-dimensional parameter space is obtained by cyclically modulating
the SSH dimerization together with a staggered on-site potential,
leading to the Rice--Mele pump
\cite{RiceMele1982,Thouless1983}. The crystal momentum and the pumping
parameter form a torus, providing a natural two-dimensional extension
of the SSH winding problem within the finite-state Bargmann framework.
This allows the topological phase accumulation to be evaluated over the
full $(k,\lambda)$ torus while retaining its cell-resolved geometric
structure.

We consider
\begin{equation}
H_{\mathrm{RM}}(k,\lambda)=
\begin{pmatrix}
\Delta_{\mathrm{RM}}(\lambda) &
t_1(\lambda)+t_2(\lambda)e^{-ik}\\
t_1(\lambda)+t_2(\lambda)e^{ik} &
-\Delta_{\mathrm{RM}}(\lambda)
\end{pmatrix},
\label{eq:rice_mele_hamiltonian}
\end{equation}
where
$t_1(\lambda)=t+\delta_{\mathrm{RM}}(\lambda)$,
$t_2(\lambda)=t-\delta_{\mathrm{RM}}(\lambda)$,
with
$\delta_{\mathrm{RM}}(\lambda)
=
\delta_0\cos\lambda$,
$\Delta_{\mathrm{RM}}(\lambda)
=
\Delta_0\sin\lambda$,
$0\leq\lambda<2\pi$.
Here $\delta_{\mathrm{RM}}$ and $\Delta_{\mathrm{RM}}$ denote the
Rice--Mele dimerization and staggered-potential parameters,
respectively, and are distinct from the Bargmann invariant
$\Delta_3$.

Writing $H_{\mathrm{RM}}=\mathbf d\cdot\boldsymbol{\sigma}$ gives
$\mathbf d(k,\lambda)=
\left(
t+\delta_{\mathrm{RM}}
+(t-\delta_{\mathrm{RM}})\cos k,\,
(t-\delta_{\mathrm{RM}})\sin k,\,
\Delta_{\mathrm{RM}}
\right)$,
with
$|\mathbf d|^2=
4t^2\cos^2\frac{k}{2}
+
4\delta_{\mathrm{RM}}^2\sin^2\frac{k}{2}
+
\Delta_{\mathrm{RM}}^2$.
A band touching requires
$k=\pi$, $\delta_{\mathrm{RM}}=0$, and $\Delta_{\mathrm{RM}}=0$.
The pumping cycle encircles this gap-closing point without passing
through it, so that the lower band remains isolated throughout the
parameter-space torus $(k,\lambda)\in T^2$.

We discretize this torus using a periodic
$N_k\times N_\lambda$ mesh,
$k_i=2\pi i/N_k$ and $\lambda_j=2\pi j/N_\lambda$, with periodic
identification in both directions. Each elementary square is divided
along a diagonal into two consistently oriented triangles. The
finite-state geometric phases and the corresponding cell-resolved
densities are then constructed directly according to the definitions
of Sec.~\ref{sec:bargmann_construction}. In particular, the two
triangle phases combine according to Eq.~\eqref{eq:square_phase_sum},
and the phase-sector cell density is obtained from
Eq.~\eqref{eq:finite_bargmann_curvature}. The corresponding amplitude-sector
density is evaluated from the modulus of the same finite-state
Bargmann construction. The global topological quantity is evaluated as the
Bargmann-evaluated Chern number from the same oriented phase
accumulation defined in Eq.~\eqref{eq:CB_general}, now over the
$(k,\lambda)$ torus.

For the representative cycle
$t=1$, $\delta_0=0.5$, and $\Delta_0=0.5$, the lower band remains
isolated throughout $T^2$. On an $N=80$ periodic mesh, all elementary
Bargmann invariants remain nonzero, with
$\min_{\triangle}|\Delta_3|=0.9875756$,
$\max_{\triangle}|\Delta_3|=0.9998675$.
The accumulated phase is $\sum_{\triangle}\gamma_\triangle=-2\pi$,
and hence, from Eq.~\eqref{eq:CB_general},
\begin{equation}
C_B=-1 .
\label{eq:rice_mele_CB_result}
\end{equation}

The global sum does not retain the distribution of the individual cell
phases. For the present torus, whose area is $4\pi^2$, the
cell-resolved density satisfies
\begin{equation}
\langle\mathcal K_B\rangle
=
\frac{1}{4\pi^2}
\sum_{\square}
\mathcal K_B\,h_kh_\lambda
=
-\frac{1}{2\pi}.
\label{eq:rice_mele_KB_average}
\end{equation}
Thus, while the Chern number fixes the total accumulated phase,
$\mathcal K_B$ retains its distribution over the two-dimensional
pumping parameter space.
The structure of $\mathcal K_B$ can be related to the geometry of the
Rice--Mele cycle. At $k=\pi$, the direct band separation is
$E_g(\lambda)
=
2\sqrt{
4\delta_{\mathrm{RM}}^2(\lambda)
+
\Delta_{\mathrm{RM}}^2(\lambda)
}$.
For the representative cycle, this gap is smallest at
$\lambda=\pi/2$ and $3\pi/2$, where
$\delta_{\mathrm{RM}}=0$ and
$\Delta_{\mathrm{RM}}=\pm\Delta_0$. These points correspond to the
closest approaches of the cycle to the underlying SSH gap-closing
configuration
$(k,\delta_{\mathrm{RM}},\Delta_{\mathrm{RM}})=(\pi,0,0)$, while the
finite staggered potential keeps the Rice--Mele cycle gapped.
The finite-resolution geometric density develops two pronounced
negative structures near the minimum-gap regions. For $N=80$, its minimum is
$\mathcal K_B^{\mathrm{min}}=-1.9956001$ at approximately
$(k,\lambda)=(3.0631,\,4.6338)$,
with a symmetry-related feature near
$(3.0631,\,1.6494)$. Both lie close to the minimum-gap regions around
$k=\pi$. The small displacement from the exact minimum-gap points is
consistent with the finite-cell definition of $\mathcal K_B$: the
quantity characterizes geometric phase accumulated over a finite cell,
rather than a pointwise continuum quantity. The observed correspondence
therefore identifies regions where the finite-state geometric structure
is strongly resolved, without implying a pointwise identification with
the inverse gap or with the continuum Berry curvature.

\begin{figure*}[!t]
    \centering

    \begin{minipage}[t]{0.48\textwidth}
        \centering
        \includegraphics[width=\linewidth]{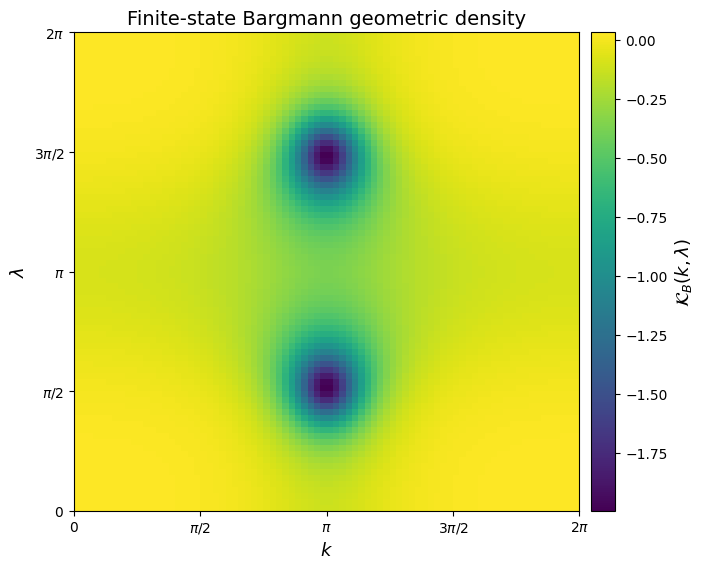}

        \vspace{2pt}
        {\small (A) $\mathcal K_B$ heat map}
    \end{minipage}
    \hfill
    \begin{minipage}[t]{0.48\textwidth}
        \centering
        \includegraphics[width=\linewidth]{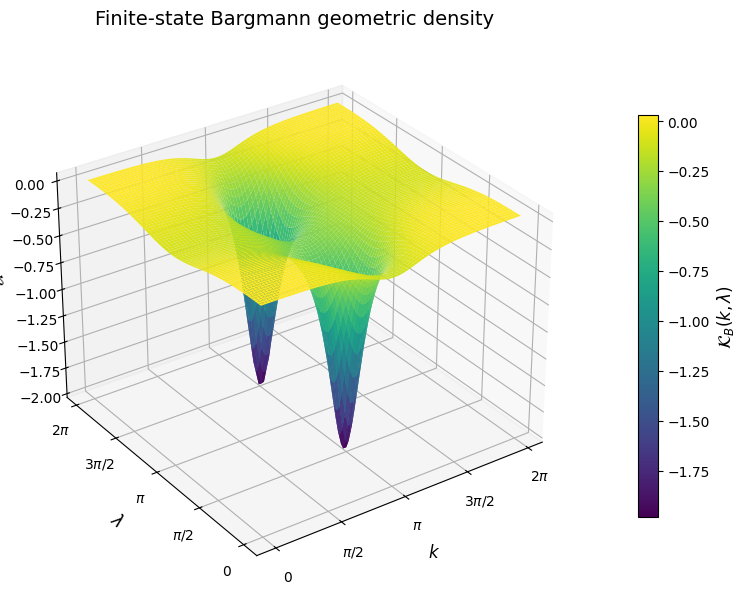}

        \vspace{2pt}
        {\small (B) Three-dimensional $\mathcal K_B$ surface}
    \end{minipage}

    \vspace{8pt}

    \begin{minipage}[t]{0.48\textwidth}
        \centering
        \includegraphics[width=\linewidth]{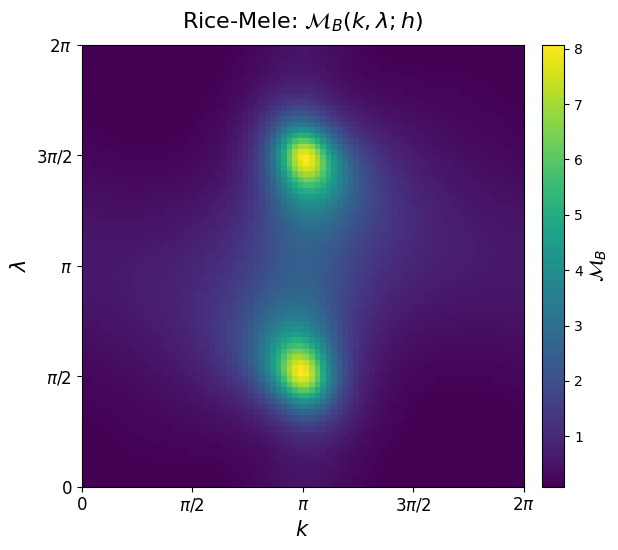}

        \vspace{2pt}
        {\small (C) $\mathcal M_B$ heat map}
    \end{minipage}
    \hfill
    \begin{minipage}[t]{0.48\textwidth}
        \centering
        \includegraphics[width=\linewidth]{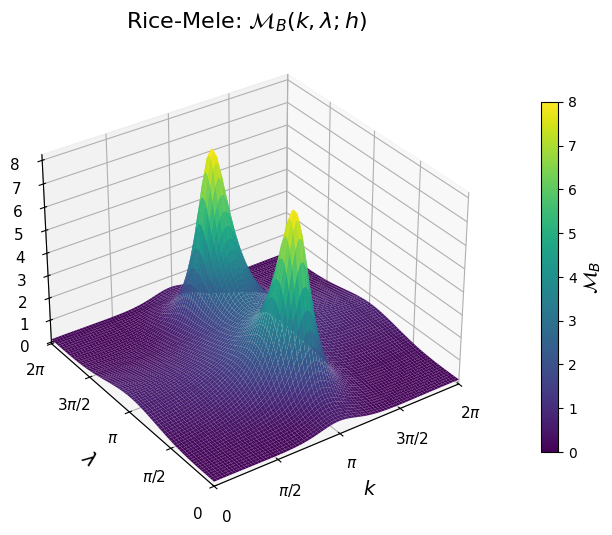}

        \vspace{2pt}
        {\small (D) Three-dimensional $\mathcal M_B$ surface}
    \end{minipage}

    \caption{
    Finite-resolution Bargmann geometric and amplitude densities for
    the Rice--Mele parameter-space torus on an $N=80$ periodic mesh.
    Panel (A) shows the phase-sector density
    $\mathcal K_B(k,\lambda)$ as a heat map, while panel (B) shows the
    same quantity as a three-dimensional surface. The two pronounced
    negative structures occur near regions of the parameter space where
    the direct band gap is smallest; their finite displacement from the
    exact minimum-gap points reflects the cell-resolved character of
    $\mathcal K_B$. Panel (C) shows the complementary amplitude-sector
    density $\mathcal M_B(k,\lambda)$ as a heat map, while panel (D)
    displays the same quantity as a three-dimensional surface.
    Together, the four panels display the phase and amplitude sectors
    obtained from the same finite-state Bargmann construction. The
    accumulated Bargmann phase gives $C_B=-1$.
    }
    \label{fig:rice_mele_local_geometry}
\end{figure*}

The positive part of the $\mathcal K_B$ distribution is comparatively
small, with $\mathcal K_B^{\mathrm{max}}=0.0312580$. The resulting
asymmetry between the pronounced negative structures and the weak
positive contribution illustrates the additional information retained
by the cell-resolved density beyond the global integer.

The local structure is resolution dependent even though the
Bargmann-evaluated Chern number is already stable on the finite meshes
considered. For
$N=6,8,10,12,16,20,24,32,40,56,80$, the minimum of $\mathcal K_B$
evolves from approximately $-0.6439$ at $N=6$ to $-1.9956$ at
$N=80$, while the global result remains $C_B=-1$. The refinement
therefore resolves progressively sharper local structure in the
finite-state geometric density while leaving the accumulated
topological phase unchanged.

The same finite-state Bargmann construction also resolves the amplitude
sector through the quantity $\mathcal M_B$ defined in
Sec.~\ref{sec:bargmann_construction}. For the present $N=80$ mesh,
$\mathcal M_B$ is strongly nonuniform over the $(k,\lambda)$ torus,
with
$\mathcal M_B^{\mathrm{min}}=0.0895773$,
$\mathcal M_B^{\mathrm{max}}=8.0716566$,
and $\langle\mathcal M_B\rangle=0.9313119$.
The corresponding heat map and three-dimensional surface are shown in
Fig.~\ref{fig:rice_mele_local_geometry}(C,D). The pronounced maximum
occurs at approximately
$(k,\lambda)=(3.0631,\,\pi/2)$,
close to the minimum-gap region
$(k,\lambda)=(\pi,\pi/2)$. This enhanced amplitude-sector structure
reflects a stronger variation of the finite-state overlap amplitudes,
and hence enhanced local distinguishability of the Bloch states, in
this region of the pumping cycle. The amplitude distribution therefore
resolves complementary information to the phase-sector density:
while $\mathcal K_B$ describes the local distribution of the geometric
phase, $\mathcal M_B$ characterizes the corresponding variation of
the overlap amplitudes. Their enhancement in nearby regions reflects
their common origin in the same finite-state quantum geometry rather
than an identification of the two quantities.

The pronounced $\mathcal M_B$ maximum also admits a simple local
interpretation. At the exact minimum-gap point
$(k,\lambda)=(\pi,\pi/2)$, the leading small-cell overlap expansion
gives $\mathcal M_B\simeq 8$. The calculated maximum
$\mathcal M_B^{\mathrm{max}}=8.0717$ occurs at
$(3.0631,\pi/2)$, one mesh step from this point, providing a direct
finite-resolution indication of the enhanced local quantum-state
variation in the minimum-gap region.

The Rice--Mele cycle extends the one-dimensional SSH construction to a
two-dimensional parameter space torus. The finite-state construction
captures the Chern number through phase accumulation and provides
complementary local information through the phase and amplitude
densities. For the representative cycle, $C_B=-1$, the phase-sector features occur near the minimum-gap regions with a finite-cell displacement that becomes less significant upon mesh refinement, while $\mathcal M_B$ develops pronounced structure in regions of strong finite-state variation. Together, the SSH and Rice--Mele examples illustrate the complementary roles of closed-loop and triangular Bargmann constructions in one- and two-dimensional parameter spaces.
\section{Summary}

We have developed and implemented a finite-state geometric framework
for band topology based on Bargmann invariants constructed directly
from quantum states on discretized parameter spaces. The third-order
Bargmann invariant provides a gauge-invariant phase associated with
three quantum states, whose oriented accumulation over a triangulated
parameter space yields the Bargmann-evaluated Chern number. Retaining
the individual cell contributions gives the finite-resolution
geometric density $\mathcal K_B$, thereby resolving how the geometric
phase is distributed across the parameter space beyond the information
contained in the global topological integer. The modulus of the same
finite-state overlap construction provides the amplitude density
$\mathcal M_B$, retaining information on the finite-state overlap
variation that is not contained in the phase accumulation alone.
Together, these local quantities preserve additional finite-resolution
information about the underlying quantum-state geometry that is
discarded when only the global accumulated phase is retained. The
continuum Berry connection and curvature are used to establish the
correspondence of these finite-state quantities with the conventional
differential-geometric description, but are not required as inputs to
the Bargmann evaluation.

The construction is demonstrated for several complementary lattice
models. In the Qi--Wu--Zhang model, the accumulated Bargmann phase
reproduces the corresponding Chern numbers in the gapped parameter
regimes. The finite-state admissibility measure also develops
characteristic features near the critical parameters, reflecting the
changing overlap structure of the Bloch states. In the $1/3$-flux
Hofstadter model, the middle band gives $C_B=-2$ on all admissible
finite meshes considered, while the local density $\mathcal K_B$
exhibits strongly nonuniform structure across the magnetic Brillouin
zone. The corresponding $\mathcal M_B$ provides additional information
on the finite-state variation of the overlap amplitudes. The finite-state
and continuum descriptions can be considered together to interpret the
geometric structure encoded in the discrete-state construction.

The $1/4$-flux Hofstadter model extends the construction to a
multiband setting in which the two central bands touch internally.
Although the individual central bands are not globally isolated, the
rank-one Bargmann invariant can still be examined in the specified
local construction near the touching. Globally, the isolated
two-dimensional central subspace is treated using overlap matrices
between local frames, leading to a gauge-invariant rank-$2$ Bargmann
invariant and the composite-subspace Chern number $C_{23}=-2$ on the
admissible finite meshes considered. The corresponding rank-$2$ phase
distribution further resolves how the composite-subspace geometric
phase is distributed across the magnetic Brillouin zone. This
demonstrates that the finite-state construction can be extended beyond
globally isolated individual bands to an isolated composite subspace,
while retaining local information about the distribution of its
geometric phase.

The SSH and Rice--Mele models demonstrate the extension of the same
finite-state overlap language to one-dimensional closed cycles and
two-dimensional pumping parameter space, respectively. In the SSH
case, the closed Bargmann product captures the phase winding around
the Brillouin zone, while the Rice--Mele cycle gives $C_B=-1$ through
the accumulated finite-state phase. The corresponding $\mathcal K_B$
and $\mathcal M_B$ retain additional local information in the
two-dimensional parameter space, including pronounced finite-resolution
structure near the minimum-gap regions. These quantities therefore
retain aspects of the state-overlap geometry that are compressed when
only the global accumulated phase is retained.

Taken together, these results establish Bargmann invariants as a
finite-state geometric language for extracting global topology
together with local finite-resolution geometric information directly
from quantum-state overlaps. The framework is not restricted to
isolated rank-one bands: it extends naturally to closed finite loops
and to isolated composite subspaces through the corresponding overlap
constructions, as demonstrated by the SSH, Rice--Mele, and
$1/4$-flux Hofstadter models. In this way, the same finite-state
framework connects global topological invariants, local finite-resolution
quantum geometry, and composite-subspace topology within a common
description based directly on quantum-state overlaps, without requiring
continuum geometric quantities as inputs to the underlying finite-state
evaluation.
\section*{Acknowledgements}

We gratefully acknowledge Utpal
Sarkar and Diptiman Sen for valuable discussions, insightful comments, and helpful
suggestions.


\begin{thebibliography}{99}

\bibitem{Berry1984}
M. V. Berry,
``Quantal phase factors accompanying adiabatic changes,''
Proc. R. Soc. A \textbf{392}, 45--57 (1984).
\href{https://doi.org/10.1098/rspa.1984.0023}{doi:10.1098/rspa.1984.0023}.

\bibitem{Thouless1982}
D. J. Thouless, M. Kohmoto, M. P. Nightingale, and M. den Nijs,
``Quantized Hall conductance in a two-dimensional periodic potential,''
Phys. Rev. Lett. \textbf{49}, 405--408 (1982).
\href{https://doi.org/10.1103/PhysRevLett.49.405}{doi:10.1103/PhysRevLett.49.405}.

\bibitem{Kohmoto1985}
M. Kohmoto,
``Topological invariant and the quantization of the Hall conductance,''
Ann. Phys. \textbf{160}, 343--354 (1985).
\href{https://doi.org/10.1016/0003-4916(85)90148-4}{doi:10.1016/0003-4916(85)90148-4}.

\bibitem{NandyTaraphderTewari2018}
S. Nandy, A. Taraphder, and S. Tewari,
``Berry phase theory of planar Hall effect in topological insulators,''
Sci. Rep. \textbf{8}, 14983 (2018).
\href{https://doi.org/10.1038/s41598-018-33258-5}{doi:10.1038/s41598-018-33258-5}.

\bibitem{Hofstadter1976}
D. R. Hofstadter,
``Energy levels and wave functions of Bloch electrons in rational and irrational magnetic fields,''
Phys. Rev. B \textbf{14}, 2239--2249 (1976).
\href{https://doi.org/10.1103/PhysRevB.14.2239}{doi:10.1103/PhysRevB.14.2239}.

\bibitem{QiWuZhang2006a}
X.-L. Qi, Y.-S. Wu, and S.-C. Zhang,
``Topological quantization of the spin Hall effect in two-dimensional paramagnetic semiconductors,''
Phys. Rev. B \textbf{74}, 085308 (2006).
\href{https://doi.org/10.1103/PhysRevB.74.085308}{doi:10.1103/PhysRevB.74.085308}.

\bibitem{QiWuZhang2006b}
X.-L. Qi, Y.-S. Wu, and S.-C. Zhang,
``A general theorem relating the bulk topological number to edge states in two-dimensional insulators,''
Phys. Rev. B \textbf{74}, 045125 (2006).
\href{https://doi.org/10.1103/PhysRevB.74.045125}{doi:10.1103/PhysRevB.74.045125}.

\bibitem{AvronSeilerSimon1983}
J. E. Avron, R. Seiler, and B. Simon,
``Homotopy and quantization in condensed matter physics,''
Phys. Rev. Lett. \textbf{51}, 51--53 (1983).
\href{https://doi.org/10.1103/PhysRevLett.51.51}{doi:10.1103/PhysRevLett.51.51}.

\bibitem{WilczekZee1984}
F. Wilczek and A. Zee,
``Appearance of gauge structure in simple dynamical systems,''
Phys. Rev. Lett. \textbf{52}, 2111--2114 (1984).
\href{https://doi.org/10.1103/PhysRevLett.52.2111}{doi:10.1103/PhysRevLett.52.2111}.

\bibitem{Bargmann1964}
V. Bargmann,
``Note on Wigner's theorem on symmetry operations,''
J. Math. Phys. \textbf{5}, 862--868 (1964).
\href{https://doi.org/10.1063/1.1704188}{doi:10.1063/1.1704188}.

\bibitem{Mukunda2003}
N. Mukunda, Arvind, E. Ercolessi, G. Marmo, G. Morandi, and R. Simon,
``Bargmann invariants, null phase curves, and a theory of the geometric phase,''
Phys. Rev. A \textbf{67}, 042114 (2003).
\href{https://doi.org/10.1103/PhysRevA.67.042114}{doi:10.1103/PhysRevA.67.042114}.

\bibitem{ProvostVallee1980}
J. P. Provost and G. Vallée,
``Riemannian structure on manifolds of quantum states,''
Commun. Math. Phys. \textbf{76}, 289--301 (1980).
\href{https://doi.org/10.1007/BF02193559}{doi:10.1007/BF02193559}.

\bibitem{SangiriSarkar2023Baryogenesis}
S. Sangiri and U. Sarkar,
``Geometric phases in kaon decays and baryogenesis,''
Nucl. Phys. B \textbf{990}, 116169 (2023).
\href{https://doi.org/10.1016/j.nuclphysb.2023.116169}
{doi:10.1016/j.nuclphysb.2023.116169}.

\bibitem{Sangiri2026Bargmann}
S. Sangiri,
``Bargmann invariants and correlated geometric $\mathcal{CP}$-violating structures in neutral meson systems,''
Eur. Phys. J. C \textbf{86}, 592 (2026).
\href{https://doi.org/10.1140/epjc/s10052-026-15853-3}{doi:10.1140/epjc/s10052-026-15853-3}.

\bibitem{Sangiri2026Sequential}
S. Sangiri,
``Connected sequential Bargmann invariants and $\mathcal{CP}$-sensitive
geometric correlation structures in neutral meson systems,''
Eur. Phys. J. C \textbf{86}, 1056 (2026).
\href{https://doi.org/10.1140/epjc/s10052-026-16308-5}
{doi:10.1140/epjc/s10052-026-16308-5}.

\bibitem{SangiriTaraphder2026Weyl}
S. Sangiri and A. Taraphder,
``Weyl chirality and Kitaev topology through Bargmann invariants,''
arXiv:2609.25278 (2026).
\href{https://arxiv.org/abs/2609.25278}{arXiv:2609.25278}.

\bibitem{SangiriSarkarTaraphder2026}
S. Sangiri, U. Sarkar, and A. Taraphder,
``Geometric phases in two-level mixing: From neutral mesons to holonomic qubits and topological Majorana modes,''
arXiv:2609.06857 (2026).
\href{https://arxiv.org/abs/2609.06857}{arXiv:2609.06857}.

\bibitem{Shashidhara2017}
R. Shashidhara,
``Driven Aubry--Andr\'e--Harper Systems,''
Master's thesis, Birla Institute of Technology and Science, Pilani (2017),
under the supervision of Dr. Tapomoy Guha Sarkar.

\bibitem{YahyaviHetenyi2017}
M. Yahyavi and B. Hetényi,
``Reconstruction of the polarization distribution of the Rice-Mele model,''
Phys. Rev. A \textbf{95}, 062104 (2017).
\href{https://doi.org/10.1103/PhysRevA.95.062104}
{doi:10.1103/PhysRevA.95.062104}.

\bibitem{AvdoshkinPopov2023}
A. Avdoshkin and F. K. Popov,
``Extrinsic geometry of quantum states,''
Phys. Rev. B \textbf{107}, 245136 (2023).
\href{https://doi.org/10.1103/PhysRevB.107.245136}
{doi:10.1103/PhysRevB.107.245136}.

\bibitem{Huang2026}
S.-M. Huang,
``Quantum-State Projectors on Grassmannian: Geometry, Holonomy, and Topology,''
arXiv:2608.06777 (2026).
\href{https://arxiv.org/abs/2608.06777}
{arXiv:2608.06777}.

\bibitem{FarragMeleLow2026}
S. Farrag, E. Mele, and T. Low,
``High Order Geometric Channels for Nonlinear Transport in Bloch Bands,''
arXiv:2607.20702 (2026).
\href{https://arxiv.org/abs/2607.20702}
{arXiv:2607.20702}.

\bibitem{Reascos2023}
L.~I.~Reascos, B.~Murta, E.~F.~Galv\~ao, and J.~Fern\'andez-Rossier,
``Quantum circuits to measure scalar spin chirality,''
Phys.\ Rev.\ Research \textbf{5}, 043087 (2023).
\href{https://doi.org/10.1103/PhysRevResearch.5.043087}{doi:10.1103/PhysRevResearch.5.043087}
\bibitem{LiWagnerZhang2026}
M.-S.~Li, R.~Wagner, and L.~Zhang,
``Multistate imaginarity and coherence in qubit systems,''
Phys.\ Rev.\ A \textbf{113}, 012428 (2026).
\href{https://journals.aps.org/pra/abstract/10.1103/tpgw-v6ht}{doi:10.1103/tpgw-v6ht}

\bibitem{Fukui2005}
T. Fukui, Y. Hatsugai, and H. Suzuki,
``Chern numbers in discretized Brillouin zone: Efficient method of computing (spin) Hall conductances,''
J. Phys. Soc. Jpn. \textbf{74}, 1674--1677 (2005).
\href{https://doi.org/10.1143/JPSJ.74.1674}{doi:10.1143/JPSJ.74.1674}.

\bibitem{SSH1979}
W. P. Su, J. R. Schrieffer, and A. J. Heeger,
``Solitons in polyacetylene,''
Phys. Rev. Lett. \textbf{42}, 1698--1701 (1979).
\href{https://doi.org/10.1103/PhysRevLett.42.1698}{doi:10.1103/PhysRevLett.42.1698}.

\bibitem{RiceMele1982}
M. J. Rice and E. J. Mele,
``Elementary excitations of a linearly conjugated diatomic polymer,''
Phys. Rev. Lett. \textbf{49}, 1455--1459 (1982).
\href{https://doi.org/10.1103/PhysRevLett.49.1455}{doi:10.1103/PhysRevLett.49.1455}.

\bibitem{Thouless1983}
D. J. Thouless,
``Quantization of particle transport,''
Phys. Rev. B \textbf{27}, 6083--6087 (1983).
\href{https://doi.org/10.1103/PhysRevB.27.6083}{doi:10.1103/PhysRevB.27.6083}.

\end{thebibliography}
\end{document}